\documentclass[a4paper,12pt]{article}
\usepackage{xcolor}
\usepackage{wrapfig}
\usepackage{amsfonts}
\usepackage{amssymb}
\usepackage{amsmath}
\usepackage{bm}
\usepackage{float}
\usepackage{graphicx}
\graphicspath{ {./Figures/} }
\usepackage{ragged2e}
\usepackage[a4paper, total={6.5in, 9.5in}]{geometry}
\usepackage{setspace}
\usepackage{caption}
\usepackage{parskip}
\usepackage{romannum}
\usepackage{hyperref}
\date{}

\usepackage{textcomp}
\usepackage{gensymb}

\usepackage{lineno}

\usepackage{etoolbox}
\usepackage{ulem}
\usepackage{xcolor}
\usepackage[numbers]{natbib}

\newcommand{\be}{\begin{equation}}
\newcommand{\ee}{\end{equation}}
\newcommand{\bea}{\begin{eqnarray}}
\newcommand{\eea}{\begin{eqnarray}}

\begin{document}
This document is the Accepted Manuscript version of a Published Work that appeared in final form in Carbon, Elsevier, after peer review and technical editing by the publisher. To access the final edited and published work see \url{https://doi.org/10.1016/j.carbon.2022.08.089}, Carbon, 201, 120, 2023.

\copyright 2022, Licensed under the Creative Commons CC-BY-NC-ND.
\pagebreak
\begin{center}

\textbf{\Large{Thermal Transport in Turbostratic Multilayer Graphene}}

\end{center}
 
\begin{center}

\author\justifying{A. Mohapatra,\textsuperscript{1,2} M. S. Ramachandra Rao,\textsuperscript{2,$\dag$} and Manu Jaiswal\textsuperscript{1,$\ast$}}

\end{center}

\begin{center} 

\textsuperscript{1}Department of Physics, Indian Institute of Technology Madras, Chennai 600036, India
 \ 
 
\textsuperscript{2}Nano Functional Materials Technology Centre and Materials Science Research Centre, Department of Physics, Indian Institute of Technology Madras, Chennai 600036, India
\ 

\end{center}

\begin{center}
Email: \textsuperscript{$\ast$}{manu.jaiswal@iitm.ac.in};
\textsuperscript{$\dag$}{msrrao@iitm.ac.in}
\end{center}

\begin{abstract}
The presence of twist angles between layers of two-dimensional materials has a profound impact on their physical properties. Turbostratic multilayer graphene is a system containing a distribution of rotational stacking faults, and these interfaces also have variable twist angles. In this work, we examine the influence of turbostratic single-layer graphene content on the in-plane thermal conductivity of a defect free multilayer graphene system with low defect density. Detailed Raman mode analysis is used to quantify the content of turbostratic single-layer graphene in the system while complementing insight is obtained from selected area electron diffraction studies. Thermal transport in these systems is investigated with Raman optothermal technique supported with finite element analysis simulations. Thermal conductivity of AB-stacked graphene diminishes by a factor of 2.59 for 1\% of turbostratic single-layer graphene content, while the decrease at 19\% turbostratic content is by an order in magnitude. Thermal conductivity broadly obeys the relation, $\kappa$ $\sim$ $\exp(-F)$, where F is the fraction of turbostratic single-layer graphene content in the system.
\end{abstract}

\section{\label{sec:level0}INTRODUCTION\protect}

Rotational stacking faults have added a whole new dimension towards engineering of the physical properties of two-dimensional materials.\cite{warner2009direct, carr2017twistronics} Artificial stacking of two-dimensional materials, as well as chemical vapor deposition (CVD) under suitable conditions, results in the formation of variable rotational angles between the layers.\cite{yang2020situ, liao2020precise} The conventional stacking order in bilayer and multilayer graphene is AB-stacking. The presence of a small twist angle between two graphene layers leads to the formation of long-range Moiré superstructures and results in spatially extended electronic wavefunction with the formation of flatbands.\cite{lisi2021observation} Superconductivity was reported in magic-angle twisted bilayer graphene (t-BLG) with one of the strongest pairing strengths between electrons.\cite{cao2018unconventional} Phononic rather than electronic properties may become the driving force behind the future applications of graphene.\cite{balandin2020phononics} Thermal transport in two-dimensional materials is also influenced by the relative rotation of atomic planes. Thermal conductivity of t-BLG was found to be reduced by a factor of $\sim$ 1.4 as compared to the value for its AB-stacked bilayer graphene counterpart.\cite{li2014thermal} Acoustic phonons are the dominant heat carriers in graphene, and the effect of layer rotation on thermal transport arises from the modification of phonon dispersion relation. Theoretical studies on t-BLG have been performed on commensurate structures corresponding to angles that retain translational symmetry. The primary influence of twisting is the large reduction in the size of the Brillouin Zone (BZ) and the formation of many folded acoustic phonon branches.\cite{PhysRevB.88.035428} The significantly reduced BZ permits additional Umklapp processes with small wavevectors, whereas several momentum forbidden phonon scattering events become allowed in the presence of multiple phonon branches. The main effect on heat transport can therefore be attributed to a change in symmetry rather than to any modification to the van der Waals interactions upon relative rotation of the layers. 

The interesting problem related to the effect of rotation of atomic planes on the heat transport in graphene pertains to the system termed as turbostratic graphene. Turbostratic graphene contains a spectrum of rotation angles corresponding to the misorientation of each layer with respect to its neighbouring layers. In completely turbostratic graphene, each layer is decoupled from other layers, and thus, the system as a whole contains Raman and electronic signatures of single-layer graphene even as several hundred layers may be present. For a system with $N$ graphene-layers and allowing for the co-existence of both AB-stacking and twisted interfaces, $2^{N-1}$ distinct stacking sequences can be constructed in principle, and for each twisted interface, some arbitrary twist angle can be defined.\cite{lin2018probing} Thus for the description of thermal transport in this system with large $N$, quantification of the twisted interfaces is required. Thermal conductivity in pyrolytic graphite with turbostratic structures has been studied in the past, but notably, that system is characterized by significant amount of edge defects with a small lateral grain size of 18-28 nm.\cite{klein1964thermal} The effects of edge disorder, and layer rotations are thus simultaneously present and defects also have a strong bearing on the thermal transport.\cite{malekpour2016thermal, raja2017annealing} In general, the Raman spectra of other related systems studied in literature including multilayer turbostratic graphene are either characterized by large defect concentration or by the presence of small turbostratic fraction or both.\cite{canccado2007measuring, pimenta2007studying} Recently, CVD has emerged as a technique to synthesize highly decoupled turbostratic single-layer graphene possessing negligible defects and also having a large grain size.\cite{gupta2020twist, mogera2015highly} In this system, not only can a large degree of turbostraticity be achieved, but the low defect density ensures that the effects arising from the statistics of layer rotations alone can be studied. With this motivation, we have investigated thermal transport in turbostratic graphene with variable fraction of turbostratic single-layer graphene content in the stack.

In this work, thermal transport was studied in high-quality multilayer CVD graphene where stacking order continuously varied from being AB-type to systems with significant turbostratic single-layer graphene content. The presence of turbostratic content was further confirmed based on the presence of stacking-order dependent combination and rotational modes in the Raman spectra. Selected-area electron diffraction provided complementing information on the presence of rotational stacking faults. The fraction of turbostratic single-layer graphene was quantified using Raman spectroscopy. Thermal conductivity was estimated using Raman optothermal technique supported with Finite Element Analysis (FEA), and its dependence on the fraction of turbostratic single-layer graphene was investigated. The low defect densities in these systems permitted to unambiguously identifying the influence of layer rotations on the thermal conductivity.

 \section{\label{sec:level1}EXPERIMENTAL\protect}
 
CVD grown AB-stacked and turbostratic multilayer graphene samples grown on Ni(111) substrates were purchased from Tata Steel Ltd.  They were subsequently transferred on to SiO\textsubscript{2}/Si substrate using thermal release tape for further characterization. The topography of the obtained graphene samples was studied using Park systems NX-10 atomic force microscopy setup in tapping mode. Raman spectroscopy of the samples along with power-dependent Raman studies were carried out using Horiba LABRam HR800 UV Raman spectrometer at laser wavelength of 632 nm (1.96 eV) along with 1800 grooves/mm grating. Variable temperature Raman scattering measurements were carried out using Linkam THMS-600 temperature stage. The incident laser power was maintained below 250 $\mu$W to avoid temperature rise of the sample due to intrinsic contribution during temperature-dependent and ambient temperature measurements. Higher laser power up to 7 mW was used for Raman optothermal studies. In the context of low-intensity combinational and rotational Raman modes associated with the nature of stacking, laser of wavelength 532 nm (2.33 eV) along with 1800 grooves/mm grating was also used (WITec alpha300 R) to obtain better signal to noise ratio. Selected-area electron diffraction (SAED) was carried out using Jeol JEM-2100 transmission electron microscopy setup.  

\section{\label{sec:level2}RESULTS AND DISCUSSION\protect}

\begin{figure}[!ht]
    \includegraphics[width=16.5cm, height=15cm]{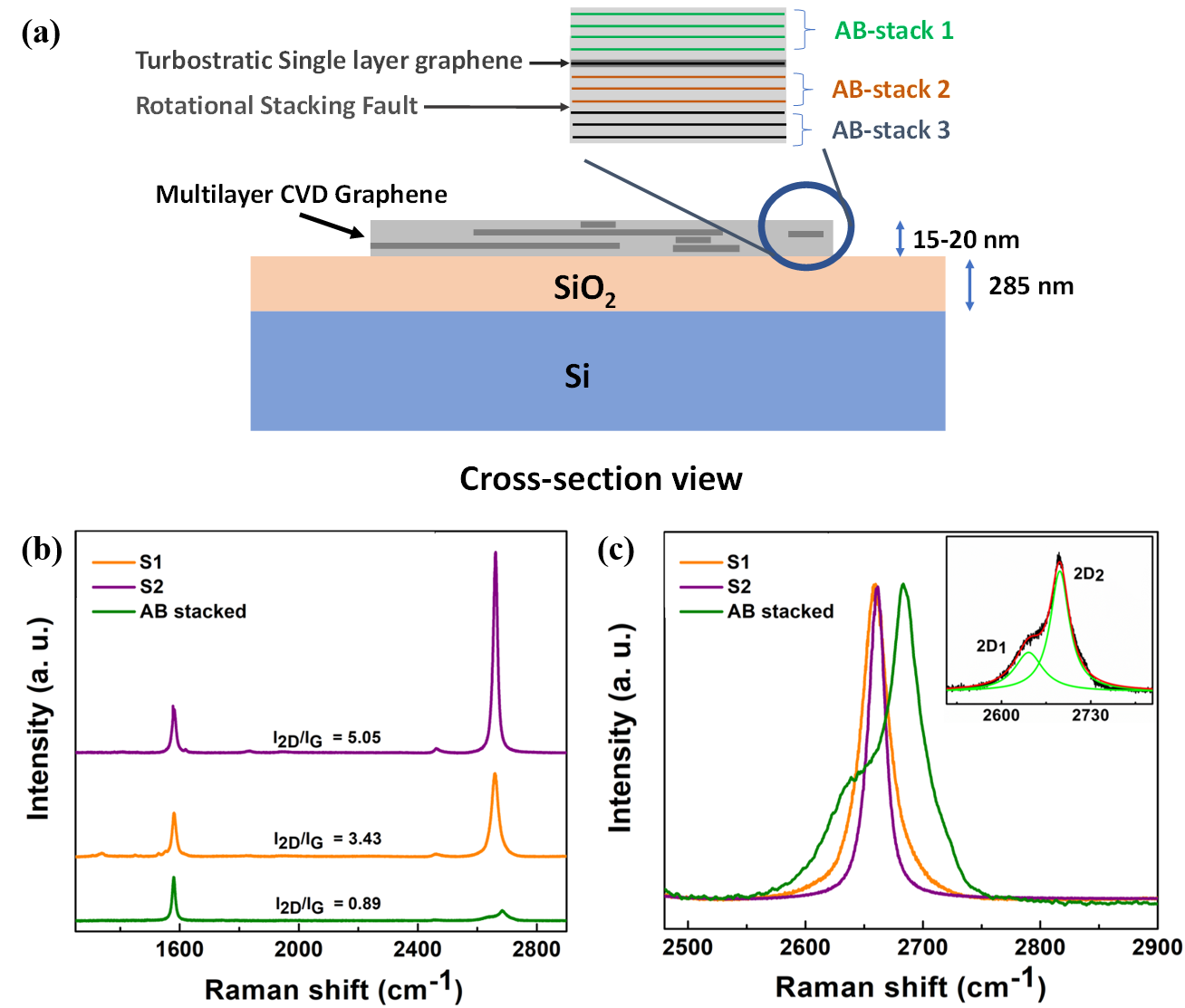}
    \
   \caption{(a) Schematic diagram showing the cross-sectional view of the multilayer CVD graphene on SiO\textsubscript{2}/Si substrate (b) Representative Raman spectra of AB-stacked and turbostratic multilayer graphene samples with increasing I\textsubscript{2D}/I\textsubscript{G} values. (c) Comparison of the 2D peak line-shape and line-width for AB-stacked and turbostratic graphene regions. The peak intensity is normalized with respect to the 2D peak intensity of turbostratic graphene. Inset shows the deconvolution of double-hump line-shape of the 2D peak in AB-stacked graphene.}
\end{figure}

Figure 1(a) shows a schematic diagram illustrating the cross-sectional view of the multilayer CVD graphene on SiO\textsubscript{2}/Si substrate. Figure 1(b) shows representative Raman spectra of multilayer turbostratic graphene, measured at two locations in comparison with the Raman spectrum of AB-stacked multilayer graphene. There are two characteristic peaks in the Raman spectrum of graphene. The G peak which corresponds to the in-plane vibrations of the lattice ascribed to E\textsubscript{2g} irreducible representation. The 2D peak corresponds to the double resonance process and is therefore sensitive to the layer number.\cite{malard2009raman} Importantly the defect peak is negligibly small in all samples confirming the high quality of the samples, see supporting information figure S\Romannum{1}. The data presented in figure 1(b) is normalized in intensity with respect to the G peak intensity of AB-stacked graphene. The Raman spectrum of AB-stacked graphene (data in olive) shows a G peak at $\sim$ 1580.12 cm\textsuperscript{-1} with full-width at half-maximum (FWHM) $\sim$ 12.67 cm\textsuperscript{-1}. The 2D peak is a convolution of two peaks with 2D\textsubscript{1} peak at $\sim$ 2717.37 cm\textsuperscript{-1} with FWHM $\sim$ 35.19 cm\textsuperscript{-1} and 2D\textsubscript{2} peak at $\sim$ 2675.92 cm\textsuperscript{-1} with FWHM $\sim$ 47.44 cm\textsuperscript{-1}(see inset of figure 1(c). It has an area integrated I\textsubscript{2D}/I\textsubscript{G} ratio of $\sim$ 0.89, and this low value, together with the double-hump line-shape is characteristic of graphitic multilayers\cite{malard2009raman} (For thickness estimates, see AFM data in supporting information figure S\Romannum{2}). In contrast, the Raman spectra of multilayer graphene with turbostratic content shows large and tuneable I\textsubscript{2D}/I\textsubscript{G} with representative values of $\sim$ 3.43 and $\sim$ 5.05 for samples S1 and S2 respectively. For sample S1 (data in orange), the G peak is observed at $\sim$ 1583.23 cm\textsuperscript{-1} with FWHM of $\sim$ 14.97 cm\textsuperscript{-1} and the 2D peak is observed at $\sim$ 2701.84 cm\textsuperscript{-1} with FWHM of $\sim$ 24.57 cm\textsuperscript{-1}. The corresponding values for sample S2 (data in purple) are G peak centered at $\sim$ 1581.07 cm\textsuperscript{-1} with FWHM of $\sim$ 15.02 cm\textsuperscript{-1} and the 2D peak is observed to be at $\sim$ 2693.80 cm\textsuperscript{-1} with FWHM of $\sim$ 18.15 cm\textsuperscript{-1}. The obtained values of $\omega$\textsubscript{G}, $\omega$\textsubscript{2D}, $\Gamma$\textsubscript{G} and $\Gamma$\textsubscript{2D} for samples S1 and S2 match well with the corresponding literature values for pristine suspended single-layer graphene.\cite{berciaud2009probing, das2008monitoring} Besides the large values for I\textsubscript{2D}/I\textsubscript{G}, the important observation is that the 2D peak appears as a single Lorentzian with narrow FWHM (see figure 1(c)), which is a characteristic signature of 2D peak of single-layer graphene, associated with the absence of multiple sub-bands in the electronic energy spectrum.\cite{malard2009raman} These observations suggest that the graphene layers are electronically decoupled with each other due to the absence of stacking order, as already discussed in the literature.\cite{gupta2020twist, mogera2015highly} The individual layers or crystallites are rotated with respect to each other and this electronic decoupling restores the single-layer behaviour. The Raman G mode intensity is proportional to the volume of the sample illuminated by the laser beam. Both AB-stacked and twisted layers contribute to the G mode intensity proportional to their respective volume fractions present in the sample. On the other hand, the single-Lorentzian 2D mode for twisted graphene contained within the stack is significantly enhanced with respect to the G mode, while the double Lorentzian 2D mode for AB-stacked graphene is suppressed with increase in turbostratic content.\cite{gupta2020twist} Coexistence of all 3 modes with distinct frequencies and intensities in the 2D peak indicates the simultaneous presence of both AB-stacked and single-layer twisted graphene regions.  However, due to the enhancement factor discussed above, it requires less than 10\% content of single-layer twisted graphene in the film to completely override the double-hump signature arising from AB-stacked regions.\textsuperscript{15}  Thus, the deconvolution of 2D mode is useful to quantify the content of twisted regions when the same is very small. It has also been empirically established in a recent study that I\textsubscript{2D}/I\textsubscript{G} ratio provides a quantification of the turbostratic content.\textsuperscript{15} The increase in the I\textsubscript{2D}/I\textsubscript{G} ratio is associated with a simultaneous increase in the number of sub-stacks which are misoriented with reference to a dominant AB-stack as well as increase in the spread of twist angles. The highest I\textsubscript{2D}/I\textsubscript{G} ratio that can be realized in principle, as obtained by extrapolation of experimental data is 17.92, and that corresponds to all sub-stacks being twisted single-layers.\cite{gupta2020twist} I\textsubscript{2D}/I\textsubscript{G} ratio serves to quantify the extent of turbostratic content when only a single Lorentzian 2D mode is observed. For these cases, 2D mode intensity almost entirely derives from turbostratic single-layer graphene regions, while all regions contribute to the G mode (for details of Raman analysis, see supporting information figure S\Romannum{3}). Based on the above analysis, the turbostratic single-layer content for the data shown in figure 1(b) is estimated to be 19.14\%, 28.18\% and 0.07\% for samples S1, S2 and AB-stacked respectively. To summarize our key observation, the fraction of twisted interfaces of single layers significantly influences the 2D mode intensity and line-shape. To further substantiate the turbostratic nature of these samples, a detailed study of the other Raman modes together with TEM-SAED is performed, and this is presented in the subsequent sections.

\begin{figure}[!ht]
    \centering
    \includegraphics[width=16cm, height=14cm]{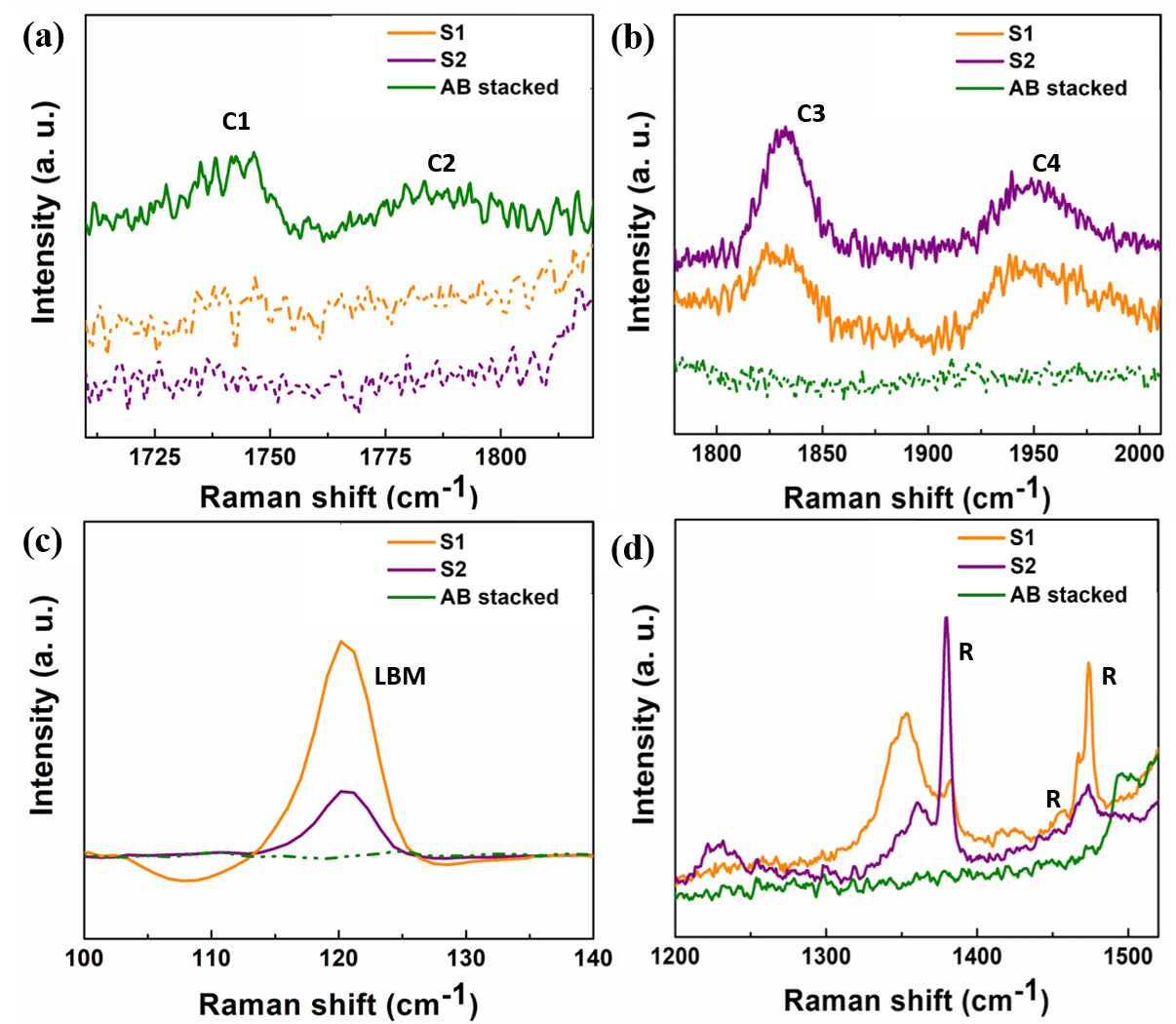}
    \caption{(a) Raman spectra showing the combination modes C1 and C2 activated in the presence of AB-stacking order, but absent in turbostratic graphene samples S1 and S2. (b) Raman spectra showing combination modes C3 and C4 activated due to loss of stacking order in turbostratic graphene samples S1 and S2, but absent in AB-stacked graphene. (c) Layer breathing mode (LBM) activated due to out of plane phonons in turbostratic graphene samples S1 and S2, but absent in AB-stacked graphene. (d) Raman spectrum showing various rotational (R) modes activated due to loss of stacking order in samples S1 and S2, but absent in AB-stacked graphene.}
\end{figure}

Several low-intensity combination and rotational Raman modes serve as identifiers for the stacking order in the system. Figure 2(a) shows two Raman active combination modes C1 and C2, which are known to be active only in AB-stacked multilayer graphene (data in olive) at $\sim$ 1740 cm\textsuperscript{-1} and 1780 cm\textsuperscript{-1} respectively. The mode C1 is shown in literature to be a combination mode activated due to the combination between LO and out of plane ZO\textsuperscript{$\prime$} phonons.\cite{lui2012observation} The total Raman shift associated with these modes is a combination of the corresponding Raman shifts associated with underlying modes of equal and opposite wave-vector. Due to this requirement, there is involvement of out of plane modes, which are active only in the \textit{presence} of stacking order. Hence these modes are absent in turbostratic systems, the data for samples S1 and S2 is also presented in figure 2(a). Similarly, the activation of Raman mode C2 can be explained by an intravalley double resonance Raman process involving the overtone mode 2ZO\textsuperscript{$\prime$}. Due to the loss of stacking order, these modes are deactivated in the case of turbostratic graphene (see data S1 and S2 in figure 2(a)). Figure 2(b) shows the Raman activated modes C3 and C4 at wavenumbers $\sim$ 1831 cm\textsuperscript{-1} and 1947 cm\textsuperscript{-1} respectively. The low-frequency mode C3 is a combination mode between iTA and LO phonons. Cong et al. showed the incident laser energy dependence of this mode shift matches very well with the combination of shifts associated with iTA and LO phonon modes.\cite{cong2011second} The higher energy mode C4 consists of two modes. The lower frequency mode in C4 is attributed to the combination between iTO and LA modes along with the high frequency mode, which is due to the combination of LO and LA modes. These modes are found to be strongly enhanced in the case of twisted bilayer graphene.\cite{cong2011second} Ramnani et al. showed the activation of modes in the range 120 cm\textsuperscript{-1} to 200 cm\textsuperscript{-1} is due to low-frequency zone folded ZA phonon modes. The peak at $\sim$ 120 cm\textsuperscript{-1} is hence attributed to the layer breathing modes activated for a fixed wavelength range due to the presence of turbostratic content in our samples.\cite{ramnani2017raman} Lin et al. also showed the presence of layer breathing modes at 120 cm\textsuperscript{-1} in multilayer graphene with stacking disorder.\cite{lin2018identifying} Figure 2(d) shows the presence of rotation (R) modes whose position and intensity are twist angle dependent, as shown in the literature. In samples S1 and S2 these R-modes are found at wavenumbers $\sim$ 1382 cm\textsuperscript{-1}, 1466 cm\textsuperscript{-1} and 1473 cm\textsuperscript{-1}. These R modes are shown to be activated due to additional momentum exchange from static potential of the superlattice structure formed in twisted multilayer graphene. The presence of this mode can be explained due to the intervalley and intravalley scattering of the phonons from the superlattice static potential. Different positions of R modes in the Raman spectrum represent different rotation angles present between layers.\cite{carozo2011raman}

\begin{figure}[!ht]
    \includegraphics[width=16.5cm, height=10.5cm]{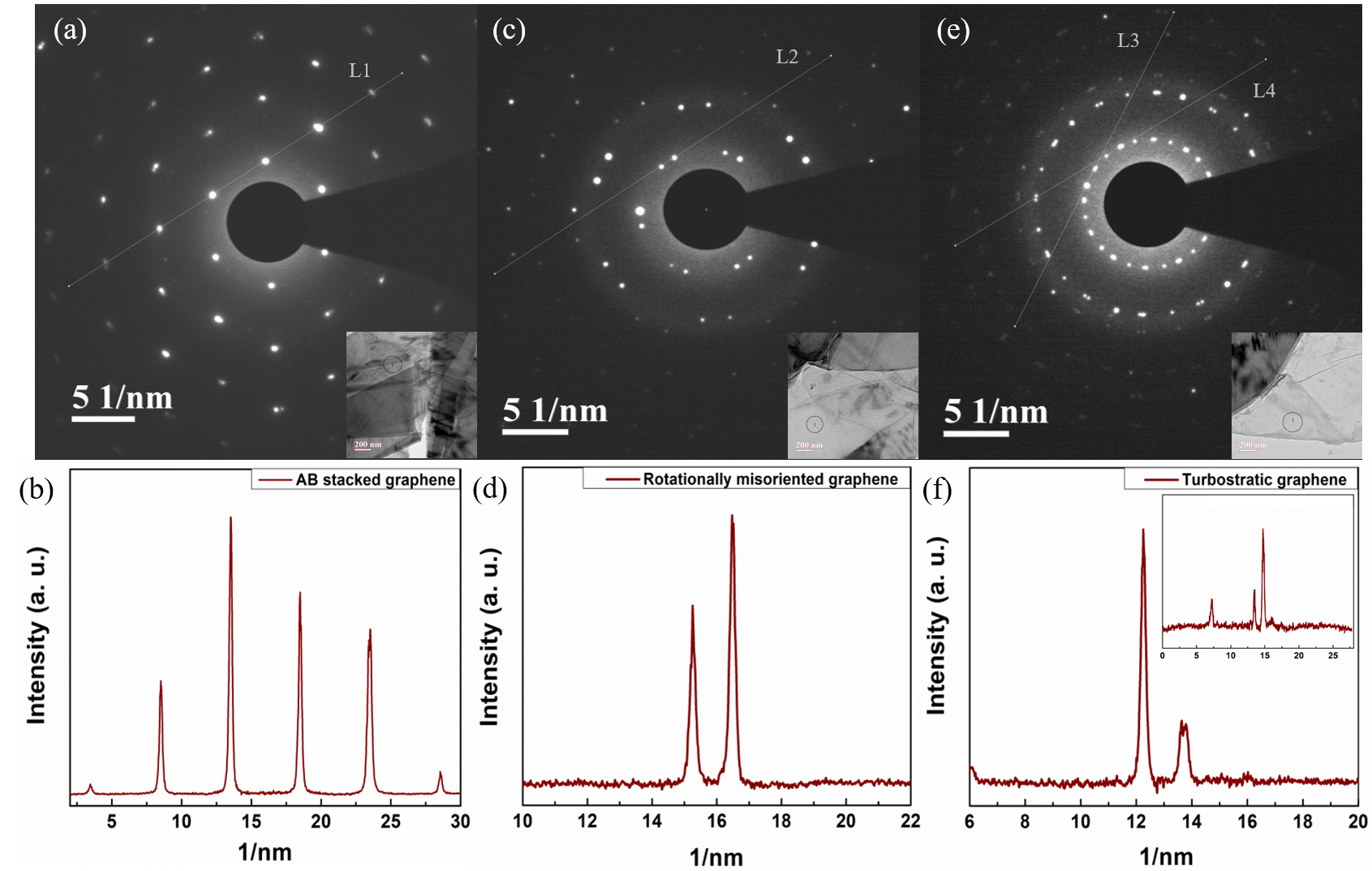}
    \caption{Electron diffraction studies: (a) SAED pattern of AB-stacked multilayer graphene. (b) Intensity profile along the line L1 in part-(a). (c) SAED of rotationally misoriented graphene. (d) Intensity profile along the line L2 shown in part-(c). (e) SAED of turbostratic graphene sample. (f) Intensity profile along one of the lines L3 in part-(e). Note, the insets at the bottom right in parts-(a), (c) and (e) show the regions marked with black circles where the corresponding diffraction data has been acquired. Inset of part-(f) shows the intensity profile along the line L4 in part-(e)}
\end{figure}

Complementing insight on the presence of rotational stacking disorder is provided by selected area electron diffraction (SAED) measurements. Figure 3(a) shows the electron diffraction pattern of a representative AB-stacked region. It consists of 6 diffraction spots which are situated at an angle of 60 degrees with respect to each other. It is due to the hexagonal D\textsubscript{6h} point group symmetry associated with the centre of graphene ring. This confirms the AB-stacking of the region. Figure 3(b) shows the intensity profile along the line L1 shown in figure 3(a). Higher intensity of the diffraction spots is representative of higher crystallinity of the system under investigation. In turbostratic systems, there are substacks or even individual layers which are not stacked parallel to the dominant AB-stack. This gives rise to multiple sets of diffraction spots in the SAED pattern. Intensity arising from each set is proportional to the volume fraction of the corresponding stack.\cite{limbu2018novel} Figure 3(c) shows the electron diffraction pattern obtained for rotationally misoriented graphene. It consists of a set of two sextets oriented at an angle of 13.12 degrees with respect to each other. Intensity of each sextet correspond to the volume fraction of a set of layers parallel to each other. In order to find out the fraction of the regions which are misoriented with respect to the dominant AB-stacked region, we considered the intensities as plotted in figure 3(d) along the line L2 as shown in figure 3(c). The spot corresponding to highest intensity is chosen as the reference spot and the ratio of the cumulative intensity of rotated spots with respect to the reference spot gives the fraction of regions which are stacked otherwise. This value is $\sim 0.37$ for the data shown in figure 3(d). Figure 3(e) shows the diffraction pattern for a sample with higher turbostratic content. It consists of many sets of diffraction spots. It shows not just one but increasingly many number of stacks which are rotated with respect to each other. The intensity of all those stacks rotated with respect to the highest intensity spot is mapped along the two straight lines (L3 and L4) as shown in figure 3(e). Intensity profile for line L3 is shown in figure 3(f) while the inset in figure 3(f) shows the intensity profile along the line L4. We obtained nearly 50 $\%$ of the stacks that are misoriented with respect to the dominant AB-stack.  The rotated sextets probed in SAED may themselves be AB-substacks which are misoriented with respect to the dominant AB-substack, while the signal from isolated single-layer twisted graphene will be quite small unless there is a preferred angle of twist. Raman mode analysis and SAED, therefore, probe quite different aspects of turbostraticity, Raman 2D mode as well as combination and rotation modes are particularly sensitive to turbostratic single-layer graphene content while SAED only reveals the extent of rotational stacking faults in the system. Nevertheless, increase in the rotational stacking disorder inferred from SAED is associated with a concomitant increase in I\textsubscript{2D}/I\textsubscript{G} ratio and also a suppression of the Raman double Lorentzian modes. 
\
\begin{center}
\begin{figure}[!ht]
    \includegraphics[width=16cm, height=13cm]{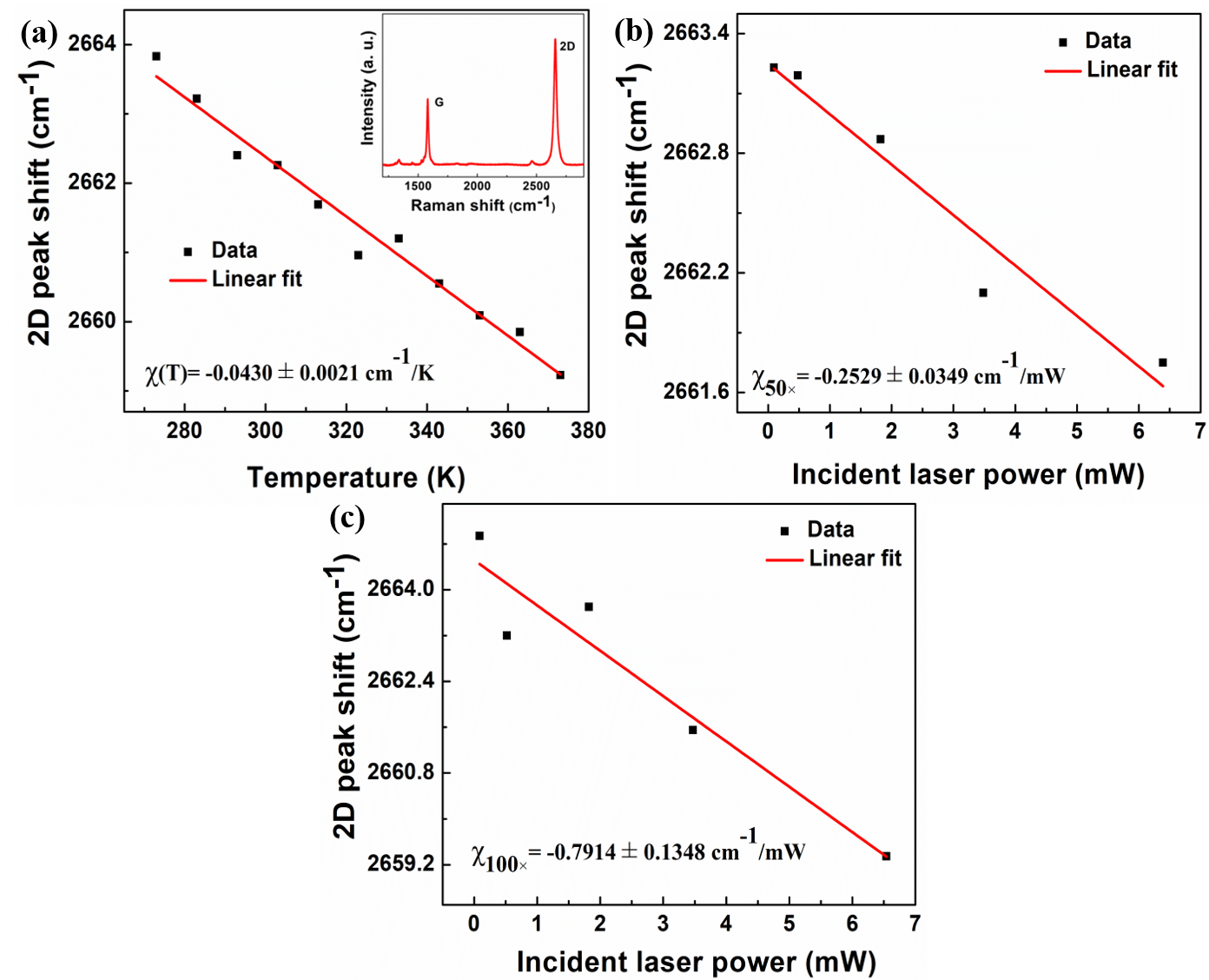}
    \caption{Raman optothermal measurements on a representative turbostratic graphene sample: (a) Temperature-dependent 2D mode shift (inset shows the Raman spectrum of turbostratic multilayer graphene). (b) Power-dependent 2D peak shift with 50$\times$ objective. (c) Power-dependent 2D peak shift with 100$\times$ objective}
\end{figure}
\end{center}
We next consider the in-plane thermal transport in these samples with variable turbostratic single-layer graphene content as quantified using Raman analysis. For Raman optothermal measurements, variable heat load is provided to the sample using the incident laser beam and the temperature variations corresponding to any given heat load is monitored using the Raman mode shifts.\cite{mohapatra2021thermal, guo2020experimental, ranjuna2022investigating, karak2021hexagonal} The Raman 2D mode shift as well as G mode shift can both equivalently serve for monitoring the temperature variations of the sample. The 2D mode has a higher resolution of temperature measurement due to higher magnitude of the associated temperature coefficient.\cite{lin2011anharmonic} For the case of turbostratic graphene, the 2D mode lineshape is given by a single Lorentzian, and the same is chosen for monitoring the temperature changes. In case of AB-stacked graphene, the 2D mode is a double hump, thus requiring deconvolution. Hence, the G mode with single Lorentzian lineshape was preferred for monitoring the temperature changes in these samples. The inset of figure 4(a) shows the Raman spectrum of turbostratic graphene with I\textsubscript{2D}/I\textsubscript{G} = 3.17 $\pm$ 0.3 and negligible defect peak.  Figure 4(a) shows the 2D mode shift as a function of temperature variation for a representative turbostratic graphene sample of thickness $\sim$ 10 nm (see supporting information figure S\Romannum{2}(d),(e) for AFM image and height profile). The obtained slope is $\chi$(T) = -0.0430 $\pm$ 0.0021 cm\textsuperscript{-1}/K. The red shift associated with the 2D mode shown in figure 4(a) arises due to the contribution from intrinsic and extrinsic factors such as thermal expansion, anharmonicity and the thermal expansion coefficient mismatch induced strain in the lattice, respectively.\cite{calizo2007temperature} Figures 4(b), (c) show the results obtained from the power dependent measurements. Figure 4(b) shows the 2D mode shift as a function of power obtained using a 50$\times$ objective. The slope from a linear fit is obtained as, $\chi$(P) = -0.2529 $\pm$ 0.0349 cm\textsuperscript{-1}/mW. Similarly, the slope obtained when using a 100$\times$ objective is $\chi$(P) = -0.7914 $\pm$ 0.1348 cm\textsuperscript{-1}/mW. $\chi$(P)(100$\times$) is higher than $\chi$(P)(50$\times$) due to higher energy density associated with the 100$\times$ objective for a given incident laser power. Results for other regions are shown in the supporting information figure S\Romannum{4}. The data discussed above provides the experimental temperature rises for different heat loads. To further obtain the thermal conductivity of the sample requires the solution of the pair of diffusion equations for the given geometry, as discussed in the following paragraph. 

The steady state temperature distribution in the film \textit{T(r)} can be obtained by solving the following heat diffusion equations, which respectively correspond to the heat flow through the thin film and through the substrate:

\begin{equation}
\kappa h \frac{1}{r}\frac{\partial}{\partial r} {\bigg(}r\frac{\partial T(r)}{\partial r} {\bigg)} - g(T(r)-T_{s}(r, 0)) + \frac{Q}{\pi r_{0}^2} e{^{-\frac {r^2}{r_{0}^2}}} = 0
\end{equation}

\begin{equation}
\kappa\textsubscript{s}\frac{1}{r}\frac{\partial}{\partial r}{\bigg(}r\frac{\partial T\textsubscript{s}(r, z)}{\partial r} {\bigg)}+\kappa\textsubscript{s}\frac{\partial^2 T\textsubscript{s}(r, z)}{\partial z^2} = 0
\end{equation}
\ \
The following boundary conditions are applied:
\begin{equation}
\frac{\partial T(r)}{\partial r}\Bigg|\textsubscript{r=0} = 0 
 \ \  
\textrm{and}
 \ \  
 T(r)|\textsubscript{r=$\infty$} = 0
\end{equation} 

\begin{equation}
\kappa\textsubscript{s}\frac{\partial T\textsubscript{s}(r, z)}{\partial z}{\bigg |}\textsubscript{z=0} = g[T(r) - T\textsubscript{s}(r, 0)]
 \ \ 
\textrm{and} 
 \ \ 
T\textsubscript{s}(r, z)|\textsubscript{z=-285 nm} = 0
\end{equation}

\begin{figure}[!ht]
    \includegraphics[width=16cm, height=9cm]{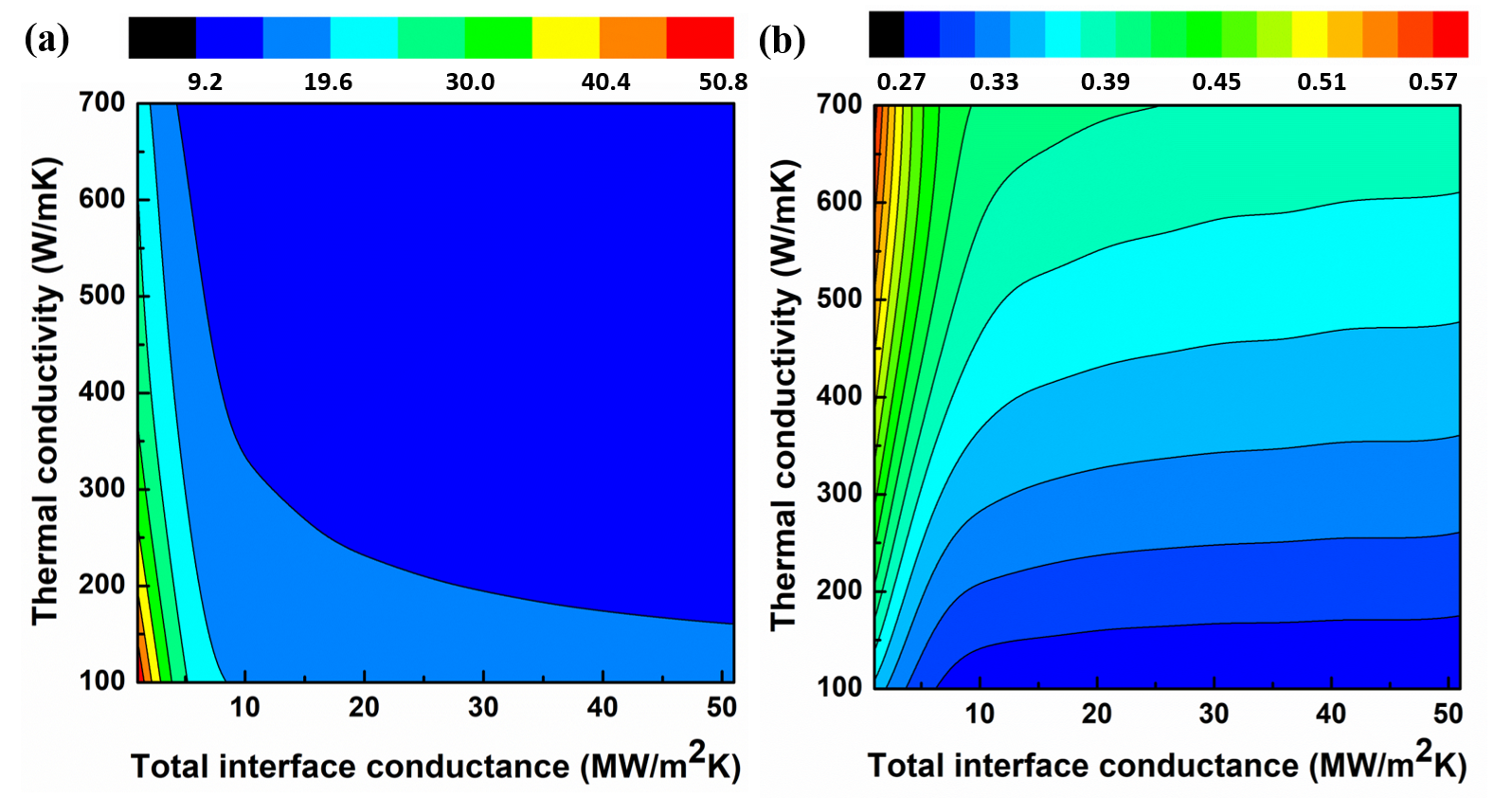}
    \caption{Finite element simulation results: (a) Temperature rise in Kelvin as a function of thermal conductivity and total interface conductance for an energy density corresponding to 100$\times$ objective. (b) Temperature rise ratio (T\textsubscript{50$\times$}/T\textsubscript{100$\times$}) as a function of thermal conductivity and total interface conductance.}
\end{figure}

In the above equations thermal conductivity of graphene is denoted as \textit{$\kappa$}, and the film thickness \textit{h}. The total interface conductance \textit{g} and \textit{T\textsubscript{s}(r, 0)} is the temperature of the substrate at the graphene substrate interface. Note that the interface between film and the substrate is taken as \textit{z = 0}. \textit{{r\textsubscript{0}}} is the incident laser spot radius. It is measured in our case using knife-edge technique and the corresponding beam spot radius values are \textit{{r\textsubscript{0}}}(100$\times$) = 1.46 $\pm$ 0.14 $\mu$m and \textit{{r\textsubscript{0}}}(50$\times$) = 3.01 $\pm$ 0.04 $\mu$m for the 100$\times$ and 50$\times$ objective lenses respectively (See supporting information figure S$\Romannum{5}$ for details). The average temperature distribution is obtained from \textit{T(r)}, following the equation given below:
 \ 
\begin{equation}
   T\textsubscript{m} =\frac{\int_{0}^{\infty} T(r)exp{\bigg(}\frac{-r\textsuperscript{2}}{r_{0}^2}{\bigg)}rdr }{\int_{0}^{\infty} exp{\bigg(}\frac{-r\textsuperscript{2}}{r_{0}^2}{\bigg)}rdr}
   \label{int}
\end{equation}

 The experiments were carried out in the ambient environment. Heat loss due to air (conductivity air $\sim$ 0.025 W/mK) is taken to be negligible.\cite{chen2011raman} We did not observe any change in the Raman mode position for Si during the measurements, and hence the SiO\textsubscript{2}/Si interface temperature was considered to be room temperature. The lateral size of the samples studied are large enough to avoid any effect on thermal properties from the boundary scattering. Furthermore, we selected regions which were nearly flat so that local curvature does not influence the thermal conductivity (see detailed discussion in supporting information S\Romannum{2}). Since thermal conductivity and total interface conductance are two independent unknown parameters, correspondingly, two different temperature rises are required to solve for them.\cite{mohapatra2021thermal} These two temperature rises were obtained by using two objective lenses with different magnifications, 50$\times$ and 100$\times$ objectives.

Finite element analysis (FEA) was carried out to evaluate the thermal conductivity and total interface conductance of the samples.\cite{judek2015high} Details of the modelling are discussed in our previous work.\cite{judek2015high, mohapatra2021thermal} In multilayer graphene, the total absorption is given by the superposition of absorptions by each individual layer, \textit{Q = n$\alpha$P}. This represents the total power absorbed by the film where $\alpha$ $\sim$ 2.3$\%$ is the absorption coefficient of single-layer graphene, $n$ represents the layer number and \textit{P} is the incident power, chosen as 1 mW for our calculations.\cite{nair2008fine} 

We now discuss the estimation of thermal conductivity and total interface conductance for samples with different turbostratic single-layer graphene content. The simulated temperature rise corresponding to 100$\times$ objective as a function of thermal conductivity and total interface conductance is shown in figure 5(a) for a multilayer graphene sample of thickness matching the experimetal one (data corresponding to the 50$\times$ objective is shown in supporting information figure S\Romannum{6}). For a given value of \textit{$\kappa$}, the temperature rise \textit{$\Delta$T} in the film decreases with increase in total interface conductance, and \textit{$\Delta$T} saturates near a minimum value showing self consistency of our calculations.

Figure 5(b) shows the temperature rise ratio T\textsubscript{50$\times$}/T\textsubscript{100$\times$} for turbostratic graphene. The temperature rise \textit{$\Delta$T} and the temperature rise ratio T\textsubscript{50$\times$}/T\textsubscript{100$\times$} are both given by contours plotted in the $\kappa$-\textit{g} plane. The measured experimental data for a sample corresponds to one of the contours for \textit{$\Delta$T} while belonging to another contour for T\textsubscript{50$\times$}/T\textsubscript{100$\times$}. The intersection between these two contours uniquely determines the values of the $\kappa$ and \textit{g}. The thermal conductivity of multilayer graphene with highest turbostratic single-layer graphene content is found to be $\sim$ 166.28 $\pm$ 52.41 W/mK and the total interface conductance $\sim$ 9.75 ± 3.53 MW/m\textsuperscript{2}K. The Raman optothermal data and FEA simulation results for AB-stacked regions are shown in the supporting information figure S\Romannum{6}. The value of thermal conductivity for AB-stacked graphene is given by $\kappa$ = 1691.29 $\pm$ 160.88 W/mK and total interface conductance, \textit{g} =  2.50 $\pm$ 0.54 MW/m${^2}$K. These values for AB-stacked graphene match very well with literature values.\cite{ghosh2010dimensional, balandin2011thermal, jang2010thickness} As discussed earlier, the I\textsubscript{2D}/I\textsubscript{G} ratio provides a measure of the turbostratic single-layer graphene content in multilayer graphene and this ratio increases monotonically with increased turbostraticity.

\begin{figure}[!ht]
    \includegraphics[width=16.5cm, height=7.1cm]{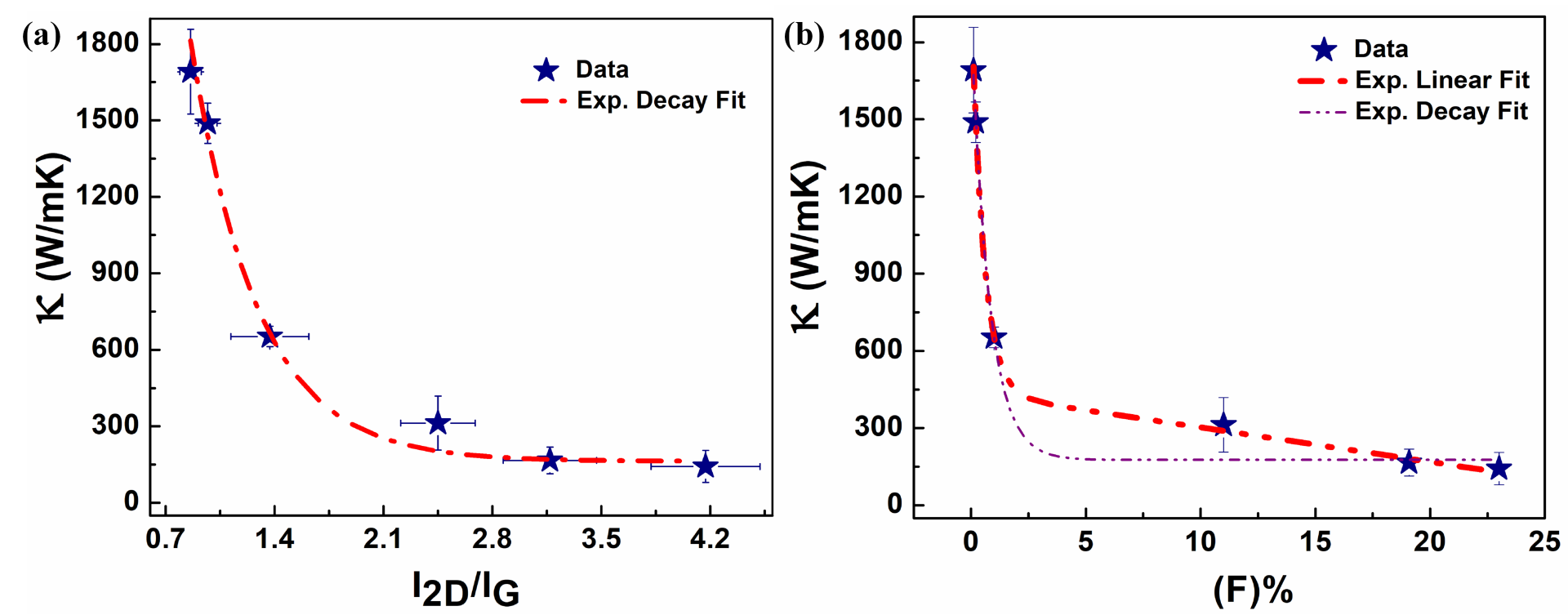}
    \caption{(a). In-plane thermal conductivity as a function of increasing  I\textsubscript{2D}/I\textsubscript{G}. (b). In-plane thermal conductivity as a function of increase in turbostratic single-layer graphene content (F\%)}
\end{figure}

A higher value of I\textsubscript{2D}/I\textsubscript{G} corresponds to both more number of twisted single-layer interfaces as well as a wider spread of rotation angles. Figure 6(a) shows a plot of the thermal conductivity as a function of I\textsubscript{2D}/I\textsubscript{G} ratio. A large decrease in thermal conductivity is found even with a slight increase in the turbostratic single-layer content in the system. Qualitatively, the decrease in thermal conductivity with increase in turbostratic single-layer graphene content can be reconciled by considering the processes described for t-BLG, which is relatively a much simpler system comprising of a single interface between two mutually rotated graphene sheets. In t-BLG, the formation of superlattice structure in the real space causes the size of the unit cell to increase. This leads to a consequent decrease in the size of the first BZ in the reciprocal space, and a large number of folded phonon branches appear in the phonon dispersion. The reduction in thermal conductivity then follows from the increase in the phase space for phonon Umklapp scattering as well as from additional scattering events which were momentum-forbidden in AB-stacked BLG.\cite{PhysRevB.88.035428} The large periodicity associated with the misoriented structure reduces the size of the BZ to such an extent that zone edge acoustic phonons can be thermally populated at room temperature.\cite{li2018commensurate} The reduction in thermal conductivity also varies with the twist angle since the size of BZ is dependent on the same. In general, the thermal conductivity of systems where transport is dominated by phonons is related to the specific heat, phonon velocity and the mean free path by the expression $\kappa$ $\sim$ C\textsubscript{v}v$\lambda$. In t-BLG, layer misorientation does not impact the specific heat and phonon velocity, whereas mean free path is greatly suppressed due to zone folding.\cite{lichenyang2019arxiv} Furthermore, the presence of several folded phonon branches can allow multiple phonon scattering processes, which were otherwise forbidden for AB-stacked graphene. For instance, in 2D SnSe sheets, a 40$\%$ reduction in thermal conductivity has been attributed to four phonon scattering processes that lead to the collapse of soft optical modes.\cite{Sun2022SnSe} In summary, additional scattering processes which were forbidden for AB-stacked graphene become allowed in the presence of turbostraticity, thereby decreasing the thermal conductivity.
The effect of rotation angle on the in-plane thermal conductivity of t-BLG was studied by Han et al. and a decrease in the thermal conductivity by $\sim$ 15\% from the highest value of $\sim$ 2100 W/mK to 1750 W/mK was observed with an increase in the twist angle by just 2$\degree$. This decrease was attributed to an increase in the phonon Umklapp scattering rates.\cite{han2021twist} Due to the presence of a large number of continuously varying twist angles and varying sequences of twisted, and Bernal stacks, theoretical investigations of thermal conductivity in turbostratic graphene are non-trivial. However, we provide an empirical relationship as given below
\ 
\ \ \ \ \ \ \ \ \ \ \ \ \ \ \ \ \ \ \ \ \ \ \ \ \ \ \ \ \ \ \ \ \ 
\
\begin{center}
$\kappa$ = $A\exp(-\beta I\textsubscript{2D}/I\textsubscript{G})$ + $\kappa$\textsubscript{0}
\end{center}
\

Where $A = 12306$, $\beta = 2.32$  and $\kappa$\textsubscript{0} = 149.4 are the fit parameters. The above relation should be considered valid for the situation where layer rotations are the primary effect in the system. To note, it has been shown in the literature that I\textsubscript{2D}/I\textsubscript{G} ratio is greatly suppressed with increase in defect cocentration in the system and the above equation clearly does not apply to systems with defects.\cite{wu2018raman} It is interesting to consider the fraction of turbostratic single-layer graphene regions required to diminish the thermal conductivity. Towards this, the Raman 2D mode data was deconvoluted to determine the intensities of 2D\textsubscript{1}, 2D\textsubscript{2} modes corresponding to AB-stacked graphene and the intensity of single-Lorentzian 2D\textsubscript{0} mode corresponding to turbostratic single-layer graphene (see supporting information figure S\Romannum{3}). The thermal conductivity is plotted as a function of the turbostratic single-layer graphene content(F) in figure 6(b). The data is empirically fitted with a similar exponentially decaying function of first order, $\kappa$(F) = $A\exp(-\beta(F))$ + $\kappa$\textsubscript{0}, and the parameters of the fitting are $A = 1699.67$, $\beta = 1.27$  and $\kappa$\textsubscript{0} = 176.24. A better fit is obtained by adding a linear correction term to the exponential function: $\kappa$(F) = $A\exp(-\beta(F))$ - $\kappa_1 F$ + $\kappa$\textsubscript{0}, and the parameters of the fitting are $A = 1535.72$, $\beta = 1.90$, $\kappa_1 = 13.39$ and $\kappa$\textsubscript{0} = 436.69. However, the exponential decay function broadly captures the dependence on turbostratic content. Twist angles might also have a bearing on the parameters of the correction terms introduced, such as the linear term. This point is discussed in detail in the paragraph below. The key finding here is that a small fraction of turbostratic single-layers (only 1\%) is sufficient to decrease the thermal conductivity by a factor of 2.59 from its maximum value of 1691.29 W/mK to 652.42 W/mK. Furthermore, 19\% turbostratic single-layer graphene is sufficient to suppress thermal conductivity by one order in magnitude. We have observed a functional dependence of $\kappa$ on a single parameter, namely the turbostratic single-layer graphene content(F). However, a concomitant increase in rotational stacking faults between AB-substack regions also happens.\cite{gupta2020twist} From a microscopic perspective both these types of interfaces contribute to decrease in $\kappa$. 

The thermal conductivity reduction in twisted BLG depends on the twist angle.\cite{li2018commensurate} Therefore, it is natural to expect that the twist angles should also determine the thermal conductivity of our multilayer graphene samples with turbostraticity. However, our samples contain a range of twist angles corresponding to turbostratic single-layer graphene. We estimated these twist angles based on the positions of the Raman R-modes \cite{jorio2013raman} and these values are tabulated in supporting information Figure S\Romannum{8}. These measurements are taken at the same spots where the thermal transport studies have been performed. Note that the maximum twist angle possible between two graphene sheets is $30^{\circ}$ when the sense of rotation, clockwise or counterclockwise, is not considered.\cite{li2018commensurate} The number of twist angles is not a measure of the degree of turbostraticity, since twist angles can potentially repeat across different vertical layers if they are favoured by the CVD growth process. Further, the line-width of R-mode could also arise from a range of closely spaced twist angles. However, the multiplicity of twist angles determines the extent of statistical averaging that needs to be considered in the heat transport.  For samples with large turbostraticity (I\textsubscript{2D}/I\textsubscript{G} = 3.34 and 5.05)  we obtained clear signals for 4 different twist angles ranging from $\sim$ $10^{\circ}$ to $\sim$ $30^{\circ}$. Therefore the decrease in thermal conductivity in our samples is based on a statistical average across these twist angles. We however note that for intermediate values of turbostraticity (I\textsubscript{2D}/I\textsubscript{G} = 1.37 and 2.44), only 2 twist angles per sample are clearly seen. In this intermediate turbostraticity region, the values of twist angles may somewhat influence the decrease in thermal conductivity since the statistical averaging process is not sufficient.

Klein et al. obtained a decrease in the in-plane thermal conductivity of polycrystalline graphite from $\sim$ 500 W/mK to $\sim$ 250 W/mK with an increase in the turbostratic content of the system. In this study the crystallite size was very small and varied from $\sim$ 18 to $\sim$ 28 nm by controlling the growth temperature between 1700 $^{\circ}$C to 2300 $^{\circ}$C.\cite{klein1964thermal} The ratio of in-plane and out-of-plane crystallite sizes, $L\textsubscript{a}/L\textsubscript{c}$ was always found to be $\sim$ 1, irrespective of the turbostratic content in the system. Hence grain boundary scattering of phonons serves as the dominant heat transport limiting factor in polycrystalline graphite.\cite{klein1964thermal} In this case it becomes impossible to separate the effect of layer rotations on thermal conductivity from the effect of edge-defect induced phonon scattering. The same is also observed for heat transport in polycrystalline t-BLG, where phonon scattering was found to be significant for grain sizes below the size of $\sim 50$ nm.\cite{limbu2017grain} Power-law dependence of thermal conductivity on the crystallite size, with $\kappa$ $\propto$ $L\textsubscript{a}\textsuperscript{1/3}$, was empirically determined for a multilayer graphene system.\cite{vlassiouk2011electrical} Furthermore, MD simulations have also confirmed the power law dependence of thermal conductivity on the grain size of the system.\cite{guo2009thermal} In related systems, such as h-BN, $\kappa$ is found to rapidly decrease with decreasing grain size for $L\textsubscript{a}$ $<$ 1 $\mu$m.\cite{ying2019tailoring} The system investigated by us has negligible defect density as evident from the Raman spectrum and the obtained average grain size L\textsubscript{a}(avg) $\sim$ 18.9 $\mu$m using Raman I\textsubscript{D}/I\textsubscript{G} ratio (see supporting information figure S\Romannum{1}). This grain size is quite large enough not to serve as a heat transport limiting factor in our case. Hence the distribution of layer rotations largely determines thermal conductivity variations in our system.

The Raman spectrum of graphene does not probe the zone center acoustic phonon modes, which are the dominant heat carriers. The linewidth of G-mode, associated with the zone center optical phonon mode, is plotted in supporting information Fig. S\Romannum{7}. The FWHM of G-mode is 12.8 cm\textsuperscript{-1} for AB-stacked graphene which increases to 15.2 cm\textsuperscript{-1} for turbostratic graphene. The linewidth ($\Gamma$) of the phonon mode is inversely proportional to the phonon lifetime, $\Gamma$ $\propto$ 1 / $\tau$. It is quite plausible that phonon lifetimes for other modes also decrease upon increase of turbostraticity.\cite{karak2021hexagonal} This may help to reconcile the experimentally observed data.

\section{\label{sec:level3}CONCLUSION\protect}

Turbostratic graphene is an interesting physical system which contains several rotational stacking faults with a wide spectrum of twist angles. While thermal transport has been studied in t-BLG, turbostratic graphene is less amenable to theoretical treatments. On the other hand, experimental efforts have been constrained by the fact that turbostratic graphite and multilayer graphene are also associated with small grain size and edge disorder. Thus, determination of the influence of layer rotations alone is difficult, unless a turbostratic multilayer graphene with low defect density is used. The in-plane thermal conductivity is found to be extremely sensitive to small fractions of turbostratic single-layer graphene content. Multilayer graphene systems are of practical importance to industry as mechanically robust heat dissipation layers. However, the presence of turbostratic content can significantly impede the heat carrying capability of the 2D systems. These results would also be useful for thermal studies on other two-dimensional systems with rotational stacking disorder.

\section*{Acknowledgements}
MJ thanks IIT Madras for Institute Research and Development Award (Early Career Level) vide project no. PH/1920/019/RFIR/008508; MSR would like to thank DST Nano – Mission funding vide project no. SR/NM/NAT/02-2005. We thank Sprint Testing Solutions for the TEM-SAED data. 

\section*{Conflicts of interest}
The authors declare no competing financial interests.

\bibliography{Akash} 

\begin{thebibliography}{10}
\expandafter\ifx\csname url\endcsname\relax
  \def\url#1{\texttt{#1}}\fi
\expandafter\ifx\csname urlprefix\endcsname\relax\def\urlprefix{URL }\fi
\expandafter\ifx\csname href\endcsname\relax
  \def\href#1#2{#2} \def\path#1{#1}\fi

\bibitem{warner2009direct}
J.~H. Warner, M.~H. R\"{u}mmeli, T.~Gemming, B.~B\"{u}chner, G.~A.~D. Briggs,
  Direct imaging of rotational stacking faults in few layer graphene, Nano
  letters 9~(1) (2009) 102--106.

\bibitem{carr2017twistronics}
S.~Carr, D.~Massatt, S.~Fang, P.~Cazeaux, M.~Luskin, E.~Kaxiras, Twistronics:
  Manipulating the electronic properties of two-dimensional layered structures
  through their twist angle, Physical Review B 95~(7) (2017) 075420.

\bibitem{yang2020situ}
Y.~Yang, J.~Li, J.~Yin, S.~Xu, C.~Mullan, T.~Taniguchi, K.~Watanabe, A.~K.
  Geim, K.~S. Novoselov, A.~Mishchenko, In situ manipulation of van der waals
  heterostructures for twistronics, Science advances 6~(49) (2020) eabd3655.

\bibitem{liao2020precise}
M.~Liao, Z.~Wei, L.~Du, Q.~Wang, J.~Tang, H.~Yu, F.~Wu, J.~Zhao, X.~Xu, B.~Han,
  et~al., Precise control of the interlayer twist angle in large scale mos2
  homostructures, Nature communications 11~(1) (2020) 1--8.

\bibitem{lisi2021observation}
S.~Lisi, X.~Lu, T.~Benschop, T.~A. de~Jong, P.~Stepanov, J.~R. Duran,
  F.~Margot, I.~Cucchi, E.~Cappelli, A.~Hunter, et~al., Observation of flat
  bands in twisted bilayer graphene, Nature Physics 17~(2) (2021) 189--193.

\bibitem{cao2018unconventional}
Y.~Cao, V.~Fatemi, S.~Fang, K.~Watanabe, T.~Taniguchi, E.~Kaxiras,
  P.~Jarillo~Herrero, Unconventional superconductivity in magic-angle graphene
  superlattices, Nature 556~(7699) (2018) 43--50.

\bibitem{balandin2020phononics}
A.~A. Balandin, Phononics of graphene and related materials, ACS Nano 14~(5)
  (2020) 5170--5178.

\bibitem{li2014thermal}
H.~Li, H.~Ying, X.~Chen, D.~L. Nika, A.~I. Cocemasov, W.~Cai, A.~A. Balandin,
  S.~Chen, Thermal conductivity of twisted bilayer graphene, Nanoscale 6~(22)
  (2014) 13402--13408.

\bibitem{PhysRevB.88.035428}
A.~I. Cocemasov, D.~L. Nika, A.~A. Balandin, Phonons in twisted bilayer
  graphene, Phys. Rev. B 88 (2013) 035428.

\bibitem{lin2018probing}
M.~L. Lin, J.~B. Wu, X.~L. Liu, P.~H. Tan, Probing the shear and layer
  breathing modes in multilayer graphene by raman spectroscopy, Journal of
  Raman Spectroscopy 49~(1) (2018) 19--30.

\bibitem{klein1964thermal}
C.~Klein, M.~Holland, Thermal conductivity of pyrolytic graphite at low
  temperatures. i. turbostratic structures, Physical Review 136~(2A) (1964)
  A575.

\bibitem{malekpour2016thermal}
H.~Malekpour, P.~Ramnani, S.~Srinivasan, G.~Balasubramanian, D.~L. Nika,
  A.~Mulchandani, R.~K. Lake, A.~A. Balandin, Thermal conductivity of graphene
  with defects induced by electron beam irradiation, Nanoscale 8~(30) (2016)
  14608--14616.

\bibitem{raja2017annealing}
S.~N. Raja, D.~Osenberg, K.~Choi, H.~G. Park, D.~Poulikakos, Annealing and
  polycrystallinity effects on the thermal conductivity of supported cvd
  graphene monolayers, Nanoscale 9~(40) (2017) 15515--15524.

\bibitem{canccado2007measuring}
L.~Can{\c{c}}ado, A.~Jorio, M.~Pimenta, Measuring the absolute raman cross
  section of nanographites as a function of laser energy and crystallite size,
  Physical Review B 76~(6) (2007) 064304.

\bibitem{pimenta2007studying}
M.~Pimenta, G.~Dresselhaus, M.~S. Dresselhaus, L.~Cancado, A.~Jorio, R.~Saito,
  Studying disorder in graphite-based systems by raman spectroscopy, Physical
  chemistry chemical physics 9~(11) (2007) 1276--1290.

\bibitem{gupta2020twist}
N.~Gupta, S.~Walia, U.~Mogera, G.~U. Kulkarni, Twist-dependent raman and
  electron diffraction correlations in twisted multilayer graphene, The Journal
  of Physical Chemistry Letters 11~(8) (2020) 2797--2803.

\bibitem{mogera2015highly}
U.~Mogera, R.~Dhanya, R.~Pujar, C.~Narayana, G.~U. Kulkarni, Highly decoupled
  graphene multilayers: turbostraticity at its best, The journal of physical
  chemistry letters 6~(21) (2015) 4437--4443.

\bibitem{malard2009raman}
L.~Malard, M.~A. Pimenta, G.~Dresselhaus, M.~Dresselhaus, Raman spectroscopy in
  graphene, Physics reports 473~(5-6) (2009) 51--87.

\bibitem{berciaud2009probing}
S.~Berciaud, S.~Ryu, L.~E. Brus, T.~F. Heinz, Probing the intrinsic properties
  of exfoliated graphene: Raman spectroscopy of free-standing monolayers, Nano
  letters 9~(1) (2009) 346--352.

\bibitem{das2008monitoring}
A.~Das, S.~Pisana, B.~Chakraborty, S.~Piscanec, S.~K. Saha, U.~V. Waghmare,
  K.~S. Novoselov, H.~R. Krishnamurthy, A.~K. Geim, A.~C. Ferrari, et~al.,
  Monitoring dopants by raman scattering in an electrochemically top-gated
  graphene transistor, Nature nanotechnology 3~(4) (2008) 210--215.

\bibitem{lui2012observation}
C.~H. Lui, L.~M. Malard, S.~Kim, G.~Lantz, F.~E. Laverge, R.~Saito, T.~F.
  Heinz, Observation of layer-breathing mode vibrations in few-layer graphene
  through combination raman scattering, Nano letters 12~(11) (2012) 5539--5544.

\bibitem{cong2011second}
C.~Cong, T.~Yu, R.~Saito, G.~F. Dresselhaus, M.~S. Dresselhaus, Second-order
  overtone and combination raman modes of graphene layers in the range of 1690-
  2150 cm- 1, ACS Nano 5~(3) (2011) 1600--1605.

\bibitem{ramnani2017raman}
P.~Ramnani, M.~R. Neupane, S.~Ge, A.~A. Balandin, R.~K. Lake, A.~Mulchandani,
  Raman spectra of twisted cvd bilayer graphene, Carbon 123 (2017) 302--306.

\bibitem{lin2018identifying}
M.~L. Lin, T.~Chen, W.~Lu, Q.~H. Tan, P.~Zhao, H.~T. Wang, Y.~Xu, P.~H. Tan,
  Identifying the stacking order of multilayer graphene grown by chemical vapor
  deposition via raman spectroscopy, Journal of Raman Spectroscopy 49~(1)
  (2018) 46--53.

\bibitem{carozo2011raman}
V.~Carozo, C.~M. Almeida, E.~H. Ferreira, L.~G. Cancado, C.~A. Achete,
  A.~Jorio, Raman signature of graphene superlattices, Nano letters 11~(11)
  (2011) 4527--4534.

\bibitem{limbu2018novel}
T.~B. Limbu, J.~C. Hern\'{a}ndez, F.~Mendoza, R.~K. Katiyar, J.~J. Razink,
  V.~I. Makarov, B.~R. Weiner, G.~Morell, A novel approach to the
  layer-number-controlled and grain-size-controlled growth of high quality
  graphene for nanoelectronics, ACS Applied Nano Materials 1~(4) (2018)
  1502--1512.

\bibitem{mohapatra2021thermal}
A.~Mohapatra, S.~Das, K.~Majumdar, M.~R. Rao, M.~Jaiswal, Thermal transport
  across wrinkles in few-layer graphene stacks, Nanoscale Advances 3~(6) (2021)
  1708--1716.

\bibitem{guo2020experimental}
M.~Guo, Y.~Qian, H.~Qi, K.~Bi, Y.~Chen, Experimental measurements on the
  thermal conductivity of strained monolayer graphene, Carbon 157 (2020)
  185--190.

\bibitem{ranjuna2022investigating}
M.~Ranjuna, J.~Balakrishnan, Investigating the thermal transport in gold
  decorated graphene by opto-thermal raman technique, Nanotechnology 33~(13)
  (2022) 135706.

\bibitem{karak2021hexagonal}
S.~Karak, S.~Paul, D.~Negi, B.~Poojitha, S.~K. Srivastav, A.~Das, S.~Saha,
  Hexagonal boron nitride--graphene heterostructures with enhanced interfacial
  thermal conductance for thermal management applications, ACS Applied Nano
  Materials 4~(2) (2021) 1951--1958.

\bibitem{lin2011anharmonic}
J.~Lin, L.~Guo, Q.~Huang, Y.~Jia, K.~Li, X.~Lai, X.~Chen, Anharmonic phonon
  effects in raman spectra of unsupported vertical graphene sheets, Physical
  Review B 83~(12) (2011) 125430.

\bibitem{calizo2007temperature}
I.~Calizo, A.~Balandin, W.~Bao, F.~Miao, C.~Lau, Temperature dependence of the
  raman spectra of graphene and graphene multilayers, Nano letters 7~(9) (2007)
  2645--2649.

\bibitem{chen2011raman}
S.~Chen, A.~L. Moore, W.~Cai, J.~W. Suk, J.~An, C.~Mishra, C.~Amos, C.~W.
  Magnuson, J.~Kang, L.~Shi, et~al., Raman measurements of thermal transport in
  suspended monolayer graphene of variable sizes in vacuum and gaseous
  environments, ACS nano 5~(1) (2011) 321--328.

\bibitem{judek2015high}
J.~Judek, A.~P. Gertych, M.~{\'S}winiarski, A.~{\L}api{\'n}ska,
  A.~Du{\.z}y{\'n}ska, M.~Zdrojek, High accuracy determination of the thermal
  properties of supported 2d materials, Scientific reports 5~(1) (2015) 1--11.

\bibitem{nair2008fine}
R.~R. Nair, P.~Blake, A.~N. Grigorenko, K.~S. Novoselov, T.~J. Booth,
  T.~Stauber, N.~M. Peres, A.~K. Geim, Fine structure constant defines visual
  transparency of graphene, Science 320~(5881) (2008) 1308--1308.

\bibitem{ghosh2010dimensional}
S.~Ghosh, W.~Bao, D.~L. Nika, S.~Subrina, E.~P. Pokatilov, C.~N. Lau, A.~A.
  Balandin, Dimensional crossover of thermal transport in few-layer graphene,
  Nature materials 9~(7) (2010) 555--558.

\bibitem{balandin2011thermal}
A.~A. Balandin, Thermal properties of graphene and nanostructured carbon
  materials, Nature Materials 10~(8) (2011) 569--581.

\bibitem{jang2010thickness}
W.~Jang, Z.~Chen, W.~Bao, C.~N. Lau, C.~Dames, Thickness-dependent thermal
  conductivity of encased graphene and ultrathin graphite, Nano letters 10~(10)
  (2010) 3909--3913.

\bibitem{li2018commensurate}
C.~Li, B.~Debnath, X.~Tan, S.~Su, K.~Xu, S.~Ge, M.~R. Neupane, R.~K. Lake,
  Commensurate lattice constant dependent thermal conductivity of misoriented
  bilayer graphene, Carbon 138 (2018) 451--457.

\bibitem{lichenyang2019arxiv}
C.~Li, R.~K. Lake, \href{https://arxiv.org/abs/1911.04639}{Small angle and
  non-monotonic behavior of the thermal conductivity in twisted bilayer
  graphene} (2019).
\newblock \href {https://doi.org/10.48550/ARXIV.1911.04639}
  {\path{doi:10.48550/ARXIV.1911.04639}}.
\newline\urlprefix\url{https://arxiv.org/abs/1911.04639}

\bibitem{Sun2022SnSe}
J.~Sun, C.~Zhang, Z.~Yang, Y.~Shen, Q.~W. M.~Hu, Four-phonon scattering effect
  and two-channel thermal transport in two-dimensional paraelectric snse, ACS
  Applied Materials and Interfaces 14 (2022) 11493–11499.

\bibitem{han2021twist}
S.~Han, X.~Nie, S.~Gu, W.~Liu, L.~Chen, H.~Ying, L.~Wang, Z.~Cheng, L.~Zhao,
  S.~Chen, Twist-angle-dependent thermal conduction in single-crystalline
  bilayer graphene, Applied Physics Letters 118~(19) (2021) 193104.

\bibitem{wu2018raman}
J.~B. Wu, M.~L. Lin, X.~Cong, H.~N. Liu, P.~H. Tan, Raman spectroscopy of
  graphene-based materials and its applications in related devices, Chemical
  Society Reviews 47~(5) (2018) 1822--1873.

\bibitem{jorio2013raman}
A.~Jorio, L.~G. Can{\c{c}}ado, Raman spectroscopy of twisted bilayer graphene,
  Solid State Communications 175 (2013) 3--12.

\bibitem{limbu2017grain}
T.~B. Limbu, K.~R. Hahn, F.~Mendoza, S.~Sahoo, J.~J. Razink, R.~S. Katiyar,
  B.~R. Weiner, G.~Morell, Grain size-dependent thermal conductivity of
  polycrystalline twisted bilayer graphene, Carbon 117 (2017) 367--375.

\bibitem{vlassiouk2011electrical}
I.~Vlassiouk, S.~Smirnov, I.~Ivanov, P.~F. Fulvio, S.~Dai, H.~Meyer, M.~Chi,
  D.~Hensley, P.~Datskos, N.~V. Lavrik, Electrical and thermal conductivity of
  low temperature cvd graphene: the effect of disorder, Nanotechnology 22~(27)
  (2011) 275716.

\bibitem{guo2009thermal}
Z.~Guo, D.~Zhang, X.~G. Gong, Thermal conductivity of graphene nanoribbons,
  Applied physics letters 95~(16) (2009) 163103.

\bibitem{ying2019tailoring}
H.~Ying, A.~Moore, J.~Cui, Y.~Liu, D.~Li, S.~Han, Y.~Yao, Z.~Wang, L.~Wang,
  S.~Chen, Tailoring the thermal transport properties of monolayer hexagonal
  boron nitride by grain size engineering, 2D Materials 7~(1) (2019) 015031.

\end{thebibliography}


\begin{thebibliography}{10}


\bibitem{1} Q. H. Wang, Z. Jin, K. K. Kim, A. J. Hilmer, G. L. C. Paulus, C. J. Shih, M. H. Ham, J. D. S. Yamagishi, K. Watanabe, T. Taniguchi, J. Kong, P. J. Herrero, and M. S. Strano, Understanding and controlling the substrate effect on graphene electron-transfer chemistry via reactivity imprint lithography, Nature Chemistry, 2012, 4, 724–732

\bibitem{2} L. G. Cançado, K. Takai, T. Enoki, M. Endo, Y. A. Kim, H. Mizusaki, A. Jorio, L. N. Coelho, R. M. Paniago, and M. A. Pimenta, General equation for the determination of the crystallite size La of nanographite by Raman spectroscopy; Applied Physics Letters, 2006, 88, 163106

\bibitem{3} N. Gupta, S. Walia, U. Mogera, $\&$ G. U. Kulkarni, Twist-Dependent Raman and Electron Diffraction Correlations in Twisted Multilayer Graphene, The Journal of Physical Chemistry Letters, 2020, 11, 2797-2803

\bibitem{4} S. Karak, S. Paul, D. Negi, B. Poojitha, S. K. Srivastav, A. Das, and S. Saha, Hexagonal Boron Nitride–Graphene Heterostructures with Enhanced Interfacial Thermal Conductance for Thermal Management Applications, ACS Appl. Nano Mater. 2021, 4, 2, 1951–1958

\bibitem{5} A. Mohapatra, S. Das, K. Majumdar, M. S. R. Rao, and M. Jaiswal, Thermal transport across wrinkles in few-layer graphene stacks, Nanoscale Advances, 2021, 3(6), 1708-1716]

\bibitem{6} L. Cui , X. Du , G. Wei and Y. Feng , Thermal Conductivity of Graphene Wrinkles: A Molecular Dynamics Simulation; J. Phys. Chem. C, 2016, 120 , 23807 —23812

\end{thebibliography}
\bibliographystyle{elsarticle-num} 

\begin{center}
\textbf{\huge{Supplementary Information}}

\

\textbf{\Large{Thermal Transport in Turbostratic Multilayer Graphene}}

\end{center}
 
\begin{center}

\author\justifying{A. Mohapatra,\textsuperscript{1,2} M. S. Ramachandra Rao,\textsuperscript{2,$\dagger$} and Manu Jaiswal\textsuperscript{1,\textsuperscript{$\ast$}}}

\end{center}
 
\begin{center} 

\textsuperscript{1}Department of Physics, Indian Institute of Technology Madras, Chennai 600036, India
 \ 
 
\textsuperscript{2}Nano Functional Materials Technology Centre and Materials Science Research Centre, Department of Physics, Indian Institute of Technology Madras, Chennai 600036, India
\ 

\end{center}

\begin{center}
Email: \textsuperscript{*}{manu.jaiswal@iitm.ac.in};
\textsuperscript{$\dagger$}{msrrao@iitm.ac.in}
\end{center}

\pagebreak

\textbf{\flushleft{\large{\Romannum{1}. Determination of average crystallite size from Raman spectroscopy}}}

\renewcommand{\thefigure}{S\Romannum{1}}
\begin{figure}[h]
    \centering
    \includegraphics[width=11.5cm, height=9cm]{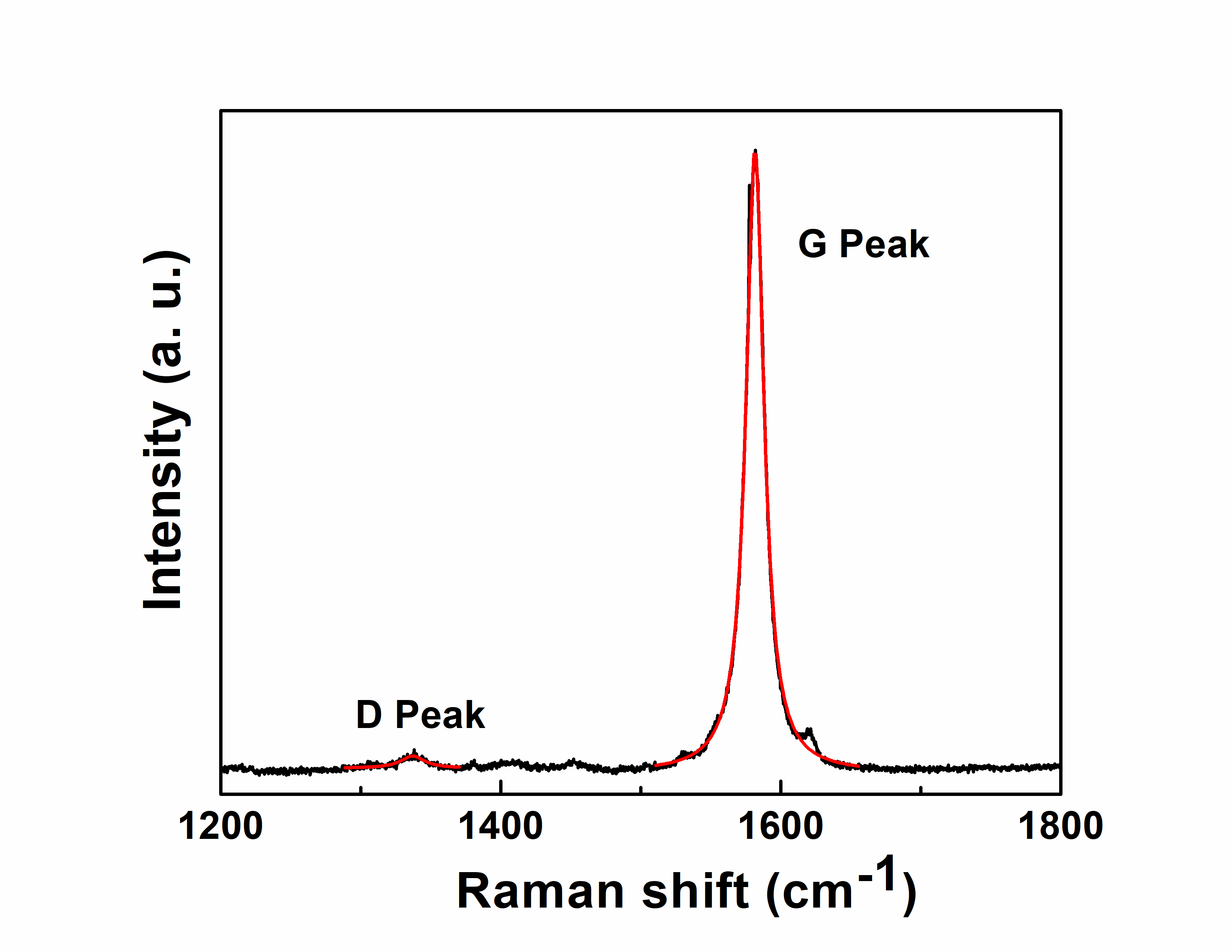}
    \caption{Averaged Raman spectrum showing the G peak and D peak intensities obtained from three turbostratic samples after adding their intensities.}
\end{figure}  

CVD graphene is associated with atomic-scale defects and grain boundaries introduced during the growth process. In all our samples, however, the defect peak intensity was very small, and the maximum value obtained across different samples had {I\textsubscript{D}/I\textsubscript{G}} {$<$} 0.09. This maximum value is also comparable to that obtained in pristine graphene systems.\textsuperscript{1} The average {I\textsubscript{D}/I\textsubscript{G}} value sampled over three turbostratic graphene regions is obtained to be {I\textsubscript{D}/I\textsubscript{G}} $\sim$ 0.008. Data in Figure S\Romannum{1} shows the Raman spectrum of turbostratic graphene showing D and G modes. Given that the I\textsubscript{D}/I\textsubscript{G} is very small, and assuming all defects are taken to be edge-type, the grain-size, {L\textsubscript{a}}, is obtained using the expression, ${L\textsubscript{a}}$ = 2.4 $\times$ 10\textsuperscript{-10} \textit{$\lambda$\textsuperscript{4} {(I\textsubscript{D}/I\textsubscript{G})\textsuperscript{-1}}}.\textsuperscript{2} The calculated value of average grain size, \textit{{L\textsubscript{a}}} $\sim$ 18.9 $\mu$m.  
\pagebreak

\textbf{\flushleft{\large{\Romannum{2}. AFM thickness profiles of AB-stacked and turbostratic graphene}}}

\renewcommand{\thefigure}{S\Romannum{2}}
\begin{figure}[h]
    \centering
    \includegraphics[width=14cm, height=12.57cm]{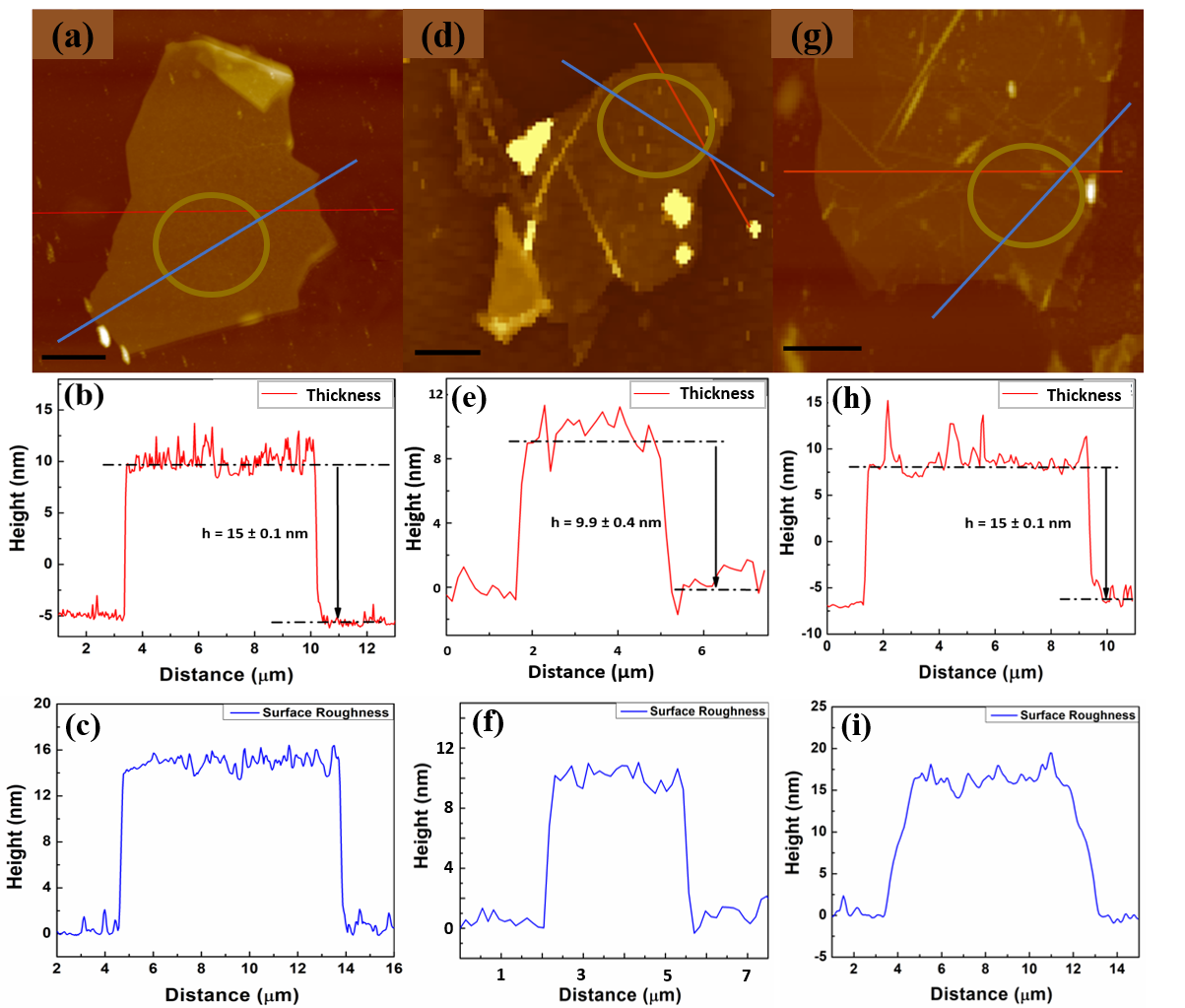}
    \caption{(a) AFM topography image of AB-stacked graphene. (b) Thickness profile of AB-stacked graphene along the red line shown in part-a. (c) Thickness profile of AB-stacked graphene along the blue line shown in part-a. (d) AFM topography image of turbostratic multi-layer graphene (sample S1 as discussed in Figure 1 of main text). (e) Thickness profile along the red line shown in part-d. (f) Thickness profile along the blue line shown in part-d. (g) AFM image of turbostratic multilayer graphene (sample S2 as discussed in Figure 1 of main text) (h) Thickness profile along the red line shown in part-g. (i) Thickness profile along the blue line shown in part-g. Regions where thermal measurements are performed are marked with dark-yellow circles in part-a,d,g. The blue line scans in these images approximately pass through the center of these regions of interest.}
\end{figure}

AFM topography measurement results on both AB-stacked and turbostratic multilayer graphene are shown in Figure S\Romannum{2}. Parts (a) and (b)  show the topography image and the thickness profile along the red line shown in part-a respectively for AB-stacked graphene. The measured thickness along the line is obtained to be $\sim$ 15 $\pm$ 0.1 nm. Part-(c) shows the roughness along the blue line in part-(a). The region marked with dark-yellow circle denote where thermal measurements were performed. Similarly the AFM topography and height profile images of turbostratic graphene is shown in part (d)-(f) (for sample S1 as discussed in the main text) and in part (g)-(i) (for sample S2 as discussed in the main text) respectively. Thickness for the sample S1 is obtained to be $\sim$ 9.9 $\pm$ 0.4 nm and for sample S2 $\sim$ 15 $\pm$ 0.1 nm respectively. The samples are originally on a Ni-substrate and transferred by exfoliation to SiO\textsubscript{2}/Si substrates in the form of films which are several microns in lateral size. Samples with and without wrinkles are generally observed in the AFM scans. In case the sample has wrinkles, then the area selected for micro-Raman optothermal measurements is taken sufficiently away from these regions so that local curvature does not influence the heat transport.

The presence of local curvature also affects thermal transport in multilayer graphene. In our previous work, we had studied the effect of wrinkles on thermal transport in graphene with rotational stacking faults.\textsuperscript{5} We had deliberately selected wrinkled locations with large curvature for the graphene film in that work. A quantitative measure of the local curvature is the ‘wrinkling level’.\textsuperscript{6} In the previous work, the wrinkle wavelength $\lambda$ $\sim$ 100 nm and amplitude A $\sim$ 9 nm resulted in a wrinkling level $\gamma$ = [A/$\lambda$] × 100$\% \sim 9\%$.

The scope of the current work is to investigate the effect of turbostraticity on the thermal transport and it is desirable that the studied samples are ultraflat. Small undulations spanning wider lateral length scales are nevertheless unavoidable on samples whose total thickness is 15-20 nm. Therefore it is necessary to identify flat regions based on the wrinkling level estimates. In the current study, we have selected only those regions with wrinkling level $<$ 0.29 $\%$ within the area corresponding to the size of the laser beam microspot. These values are obtained from the section of the blue lines enclosed within the circular regions marked in the AFM scans Fig.SII(a),(d),(g); regions where thermal measurements are performed. These regions can therefore be treated as being nearly flat. The associated curvature of any wrinkle-like feature is lower by a factor of $\sim$ 30 when compared to our previous study that specifically focused on wrinkles with large aspect ratio. 

In our previous study, we showed that the thermal conductivity increased due to formation of wrinkles with large wrinkling level ($\sim$ 9$\%$)\textsuperscript{5} and the increase was $\sim$ 35 $\%$. In the current study, with the low wrinkling level ($< 0.29\%$) for the samples, any deviation from ultraflatness is not expected to change the thermal conductivity values by more than $\sim 1\%$. This effect is negligible  when compared to the one order in magnitude change seen across our samples due to variation in turbostraticity.
 
\pagebreak 
\textbf{\flushleft{\large{\Romannum{3}. Raman 2D mode analysis for samples with increasing fraction of turbostratic single-layer graphene}}}

\renewcommand{\thefigure}{S\Romannum{3}}
\begin{figure}[h]
    \centering
    \includegraphics[width=16cm, height=10cm]{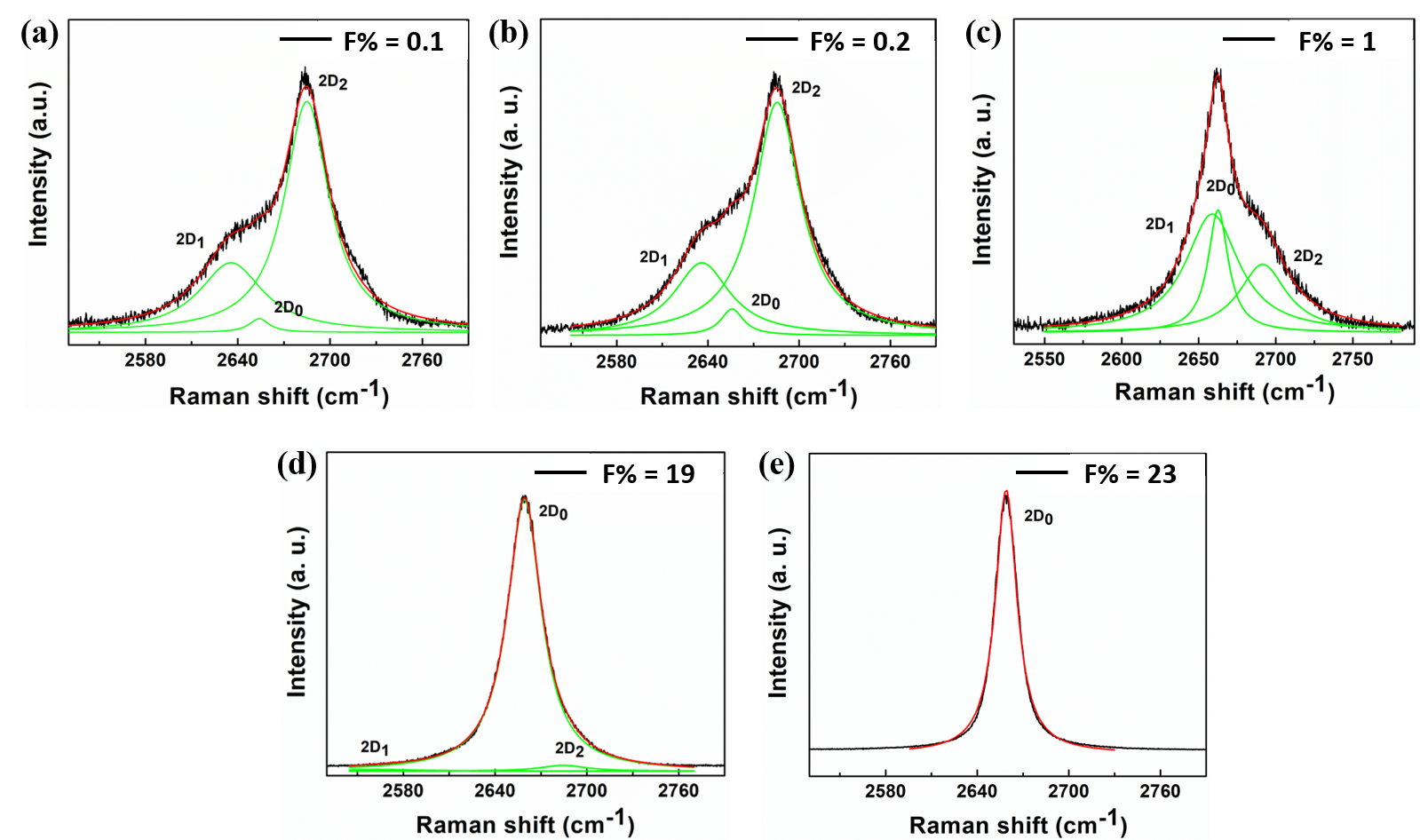}
    \caption{Raman 2D mode of multilayer graphene with an increasing order of turbostratic single-layer graphene fraction (F$\%$): (a). 0.1 $\%$ (b). 0.2 $\%$ (c). 1 $\%$ (d). 19 $\%$ (e). 23 $\%$.}
\end{figure}

In this section the fraction of turbostratic single-layer graphene contained in the multilayer graphene system is estimated using Raman mode analysis. Here F denotes the turbostratic single-layer graphene fraction and (1-F) represents the AB-stacked graphene fraction. For multilayer graphene simultaneously containing both AB-stacked and turbostratic single-layer graphene fractions, the 2D mode line-shape can be deconvoluted into three constituent modes, in general. These are denoted as $2D_1$ and $2D_2$ modes arising from the deconvolution of the double-hump 2D-peak associated with AB-stacked regions, and the single Lorentzian feature associated with turbostratic single-layer graphene, which is denoted as the $2D_0$ mode. These three modes are clearly distinct in frequency. However, deconvolution forms the basis for identifying the respective contents only when turbostratic single-layer graphene content is small (less than 10 \%).\textsuperscript{3} For larger turbostratic single-layer graphene content, the $2D_0$ mode completely dominates the signal and deconvolution is not possible. In this case,  $I\textsubscript{2D}/I\textsubscript{G}$ ratio alternatively provides the fraction of turbostratic single-layer graphene content.

Let $I_G^F$ and $I_G^{1-F}$ denote the contribution to G mode intensities due to these respective fractions. Further, $I_{2D_0}^{F}$ denotes the intensity of $2D_0$ arising from turbostratic single-layer graphene. Also,  $I_{2D_1}^{1-F}$ and $I_{2D_2}^{1-F}$ are the two deconvoluted modes intensities for the 2D peak arising from the AB-stacked fraction.

Both AB-stacked and turbostratic regions contribute identically to the G mode. It then follows,
\
$I_G^{1-F} = c(1-F)$, \ \ \ \ \ \ \ \ \  \   --------------- (1)

where $c$ is a proportionality constant giving the absolute intensity
\

$I_G^{F} = cF$ \ \ \ \ \ \ \ \ \  \   --------------- (2)
\ 

Note, the total G mode intensity arising from both types of regions is given by,
\

$I_G^{Total} = I_G^F + I_G^{1-F} = c$ --------------- (3)
\

The value of $I\textsubscript{2D}/I\textsubscript{G}$ increases with turbostratic single-layer graphene content or in other words, when more twisted single-layer graphene interfaces are present. For multilayer turbostratic graphene, the maximum value of $I\textsubscript{2D}/I\textsubscript{G}$ that can be achieved, in principle is (see main text):
\

$I_{2D_0}^{1}/I_G^{1} = 17.92$ \ \ \ \ \ \ \ \ \  \  --------- (4)

The above equation corresponds to the hypothetical situation where every interface is that of twisted single-layer graphenes. The fraction of turbostratic single-layer graphene is therefore $F = 1$ and this is indicated by the superscript provided for intensities in Eqn. (4). 
When only a fraction $F$ of the interfaces are turbostratic single-layer graphene, the $2D_0$ mode intensity decreases accordingly when compared to the maximum possible value:

$I_{2D_0}^{F} = I_{2D_0}^{1}\times F$ \ \ \ \ \ \ \ \ \  \  --------- (5)

From Eqns.(1), (2),  $I_{G}^{Total} = I_{G}^{1} = c$. Using this in Eqn.(5) gives:

$I_{2D_0}^{F}/I_{G}^{Total} = [I_{2D_0}^{1}/I_{G}^{1}]\times F$ \ \ \ \ \ \ \  \ \ \ \ \ \ --------------- (6)

Using Eqn.(4) in the RHS of Eqn.(6) then gives:

$I_{2D_0}^{F}/I_{G}^{Total} = 17.92\times F$ \ \ \ \ \ \ \  \ \ \ \ \ \ --------------- (7)

CASE-1: Eqn.(7) provides a means to estimate the turbostratic single-layer graphene fraction, $F$, when a single Lorentzian 2D peak ($2D_0$ mode) alone is observed. 
      
CASE-2: Now we consider the case when turbostratic single-layer graphene fraction is small such that deconvolution into 3 modes is possible. Our data for lowest $I\textsubscript{2D}/I\textsubscript{G}$ value of 0.86, which corresponds to AB-stacked multilayer graphene gives 
\ 
$I_{2D}^{Total, 1-F} = I_{2D_1}^{1-F} + I_{2D_2}^{1-F} = 0.86 I_G^{1-F} = 0.86c(1-F)$\ \ \ \ \ \ \ \ \  \ \ \  --------------- (8)
\

Dividing Eqn.(7) by Eqn.(8), and using Eqn.(3):

$I_{2D_0}^{F}/I_{2D}^{Total, 1-F} = 20.83\times[F/(1-F)]$  -------------- (9)
\ 

Eqn.(9) provides the turbostratic single-layer graphene content, $F$ from the ratio of the intensity of the $2D_0$ mode with the summed intensities of $2D_1$ and $2D_2$ modes. This procedure is adopted for smaller $I_{2D}/I_G$ ratios when deconvolution is possible.

Note, the above toy model is expected to give accurate results when large number of interfaces are present such that the Raman laser probes AB-stacked and turbostratic single-layer graphene regions according to their respective fractional content. When very few interfaces are present, the exact position of the interface in the stack also matters and interfaces which are closer to the top surface would provide bigger contributions to the detected signal due to light attenuation. 

Figure S\Romannum{3}(a) shows the 2D peak for AB-stacked graphene with $I\textsubscript{2D}/I\textsubscript{G} = 0.86$. It is fitted with 3 peaks as indicated by green curves. Presence of small $2D\textsubscript{0}$ peak is a signature of the simultaneous presence of both AB-stacked and turbostratic single-layer graphene content in the system. Eqn.(4) can be used in this case for the determination of $F$ fraction. It is found to be $\sim$ 0.1$\%$, which is very low. Figure S\Romannum{3}(b) shows the 2D peak for AB-stacked graphene with $I\textsubscript{2D}/I\textsubscript{G} = 0.97$. The figure illustrates that with an increase in the $I\textsubscript{2D}/I\textsubscript{G}$ value, the $2D\textsubscript{0}$ peak intensity also increases gradually. The obtained value of $F$ fraction is $\sim$ 0.2$\%$. Figure S\Romannum{3}(c) shows a further increase in the $2D\textsubscript{0}$ peak intensity value with an increase in $I\textsubscript{2D}/I\textsubscript{G}$. It finally dominates over $2D\textsubscript{1}$ and $2D\textsubscript{2}$ peak intensities with further increase in I\textsubscript{2D}/I\textsubscript{G} value. From the data in Figure S\Romannum{3}(c), the $F$ fraction is obtained to be $\sim 1\%$. It subsequently increases to $19\%$ and $23\%$ respectively for the data in parts S\Romannum{3}(d) and S\Romannum{3}(e). These data correspond to a I\textsubscript{2D}/I\textsubscript{G} value of 3.47 and 4.17 respectively.  All the data except the data in part-e are fitted with three peaks. With an increase of the turbostratic single-layer graphene content to just 4$\%$ the single Lorentzian behaviour of the 2D peak is observed.

\pagebreak
\textbf{\flushleft{\large{\Romannum{4}. Temperature and power dependent Raman optothermal measurement results for other AB-stacked and turbostratic graphene samples}}}

\renewcommand{\thefigure}{S\Romannum{4}}
\begin{figure}[h]
    \centering
    \includegraphics[width=14.5cm, height=16cm]{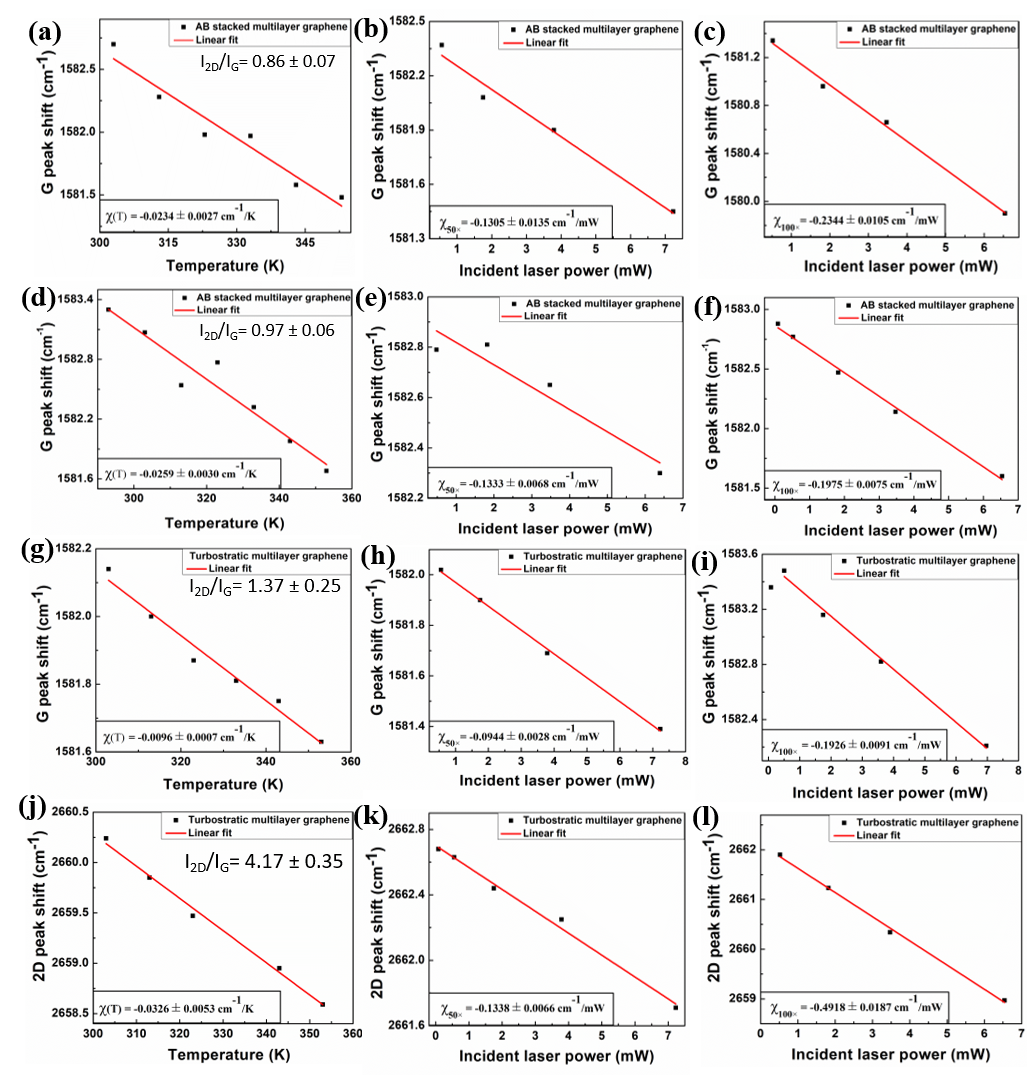}
    \caption{Temperature and power dependent (50$\times$ and 100$\times$ objectives respectively) phonon mode shifts for different I\textsubscript{2D}/I\textsubscript{G} ratio values: (a),(b),(c). I\textsubscript{2D}/I\textsubscript{G} = 0.86$\pm$0.07, (d),(e),(f).  I\textsubscript{2D}/I\textsubscript{G} = 0.97$\pm$0.06, (g),(h),(i). I\textsubscript{2D}/I\textsubscript{G} = 1.37$\pm$0.25, (j),(k),(l).
    I\textsubscript{2D}/I\textsubscript{G} = 4.17$\pm$0.35.}
\end{figure}

\pagebreak

\textbf{\flushleft{\large{\Romannum{5}. Laser spot size determination: knife-edge technique}}}

\renewcommand{\thefigure}{S\Romannum{5}}
\begin{figure}[h]
    \centering
    \includegraphics[width=12cm, height=10.28cm]{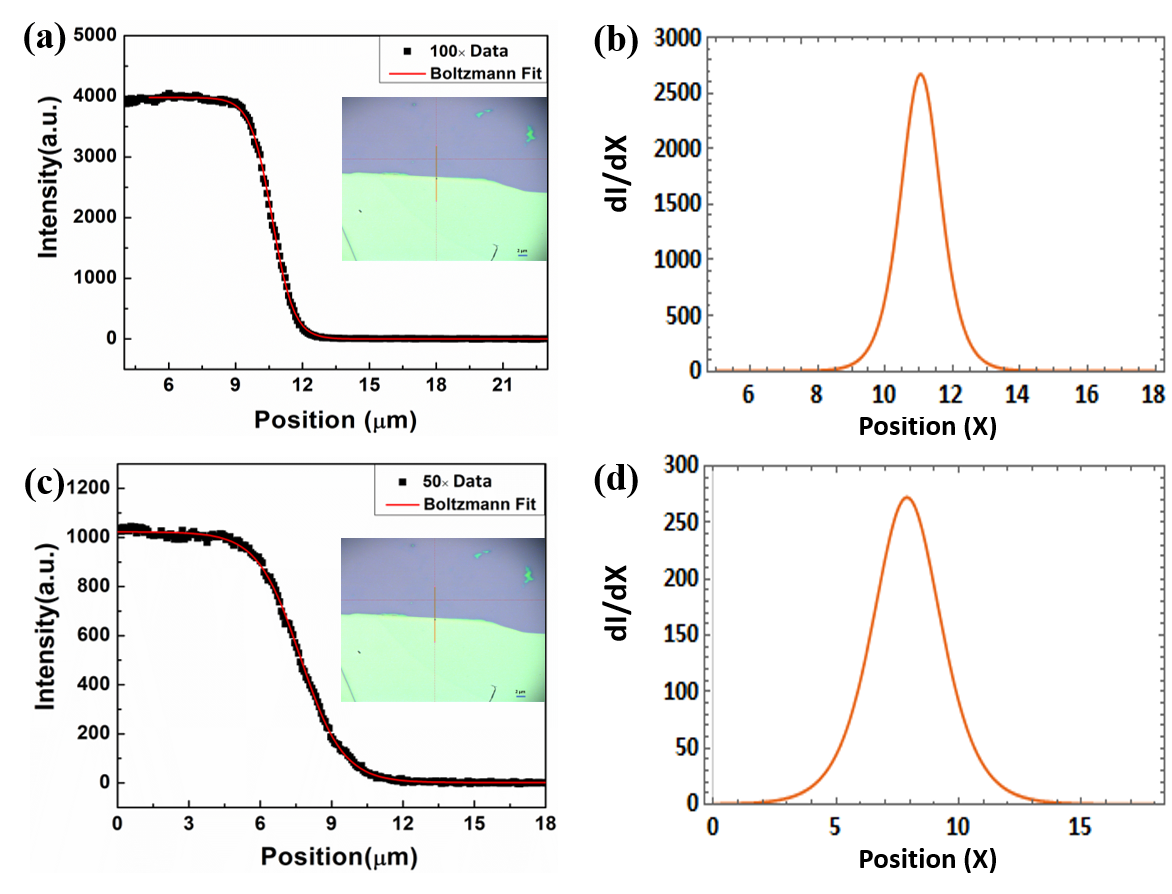}
    \caption{(a) Intensity line profile across a sharp edge obtained using 100$\times$ objective (b) Gaussian nature of the laser spot profile for 100$\times$ objective (c) Intensity line profile across a sharp edge obtained using 50$\times$ objective (d) Gaussian nature of the the laser spot profile for 50$\times$ objective.}
\end{figure}
Laser spot radius was determined using knife-edge technique. In this technique the laser is scanned across a sharp edge of thick bulk graphite exfoliated onto SiO\textsubscript{2}/Si substrate and the Raman scattered intensity of the silicon peak is measured. The advantage of using graphite edge is its atomically sharp edge as shown in the optical image in the inset of part S\Romannum{5}(a), (c) for 100$\times$ and 50$\times$ objectives respectively. The thickness of graphite is chosen such that no silicon peak is observed when the beam is measuring over the graphite region. The intensity profile is shown in Figure S\Romannum{5}(a) for 100$\times$ objective. Figure S\Romannum{5}(b) shows the differentiated intensity as a function of position. It is Gaussian in nature and the spot size can be determined from the fitting parameters. The spot size is obtained from the data where the intensity falls to 1/e\textsuperscript{2} of the peak intensity and its value is \textit{{r\textsubscript{0}}}(100$\times$) = 1.46 $\pm$ 0.14 $\mu$m. The spot size for 50$\times$ objective is obtained using a similar procedure. The intensity profile using 50$\times$ objective is shown in part S\Romannum{5}(c). The obtained differential intensity is shown in part (d). From this data, the value of the spot radius is \textit{{r\textsubscript{0}}}(50$\times$) = 3.01 $\pm$ 0.04 $\mu$m.

\pagebreak
\textbf{\flushleft{\large{\Romannum{6}. Finite element simulation results}}}

\renewcommand{\thefigure}{S\Romannum{6}}
\begin{figure}[h]
    \centering
    \includegraphics[width=13cm, height=13cm]{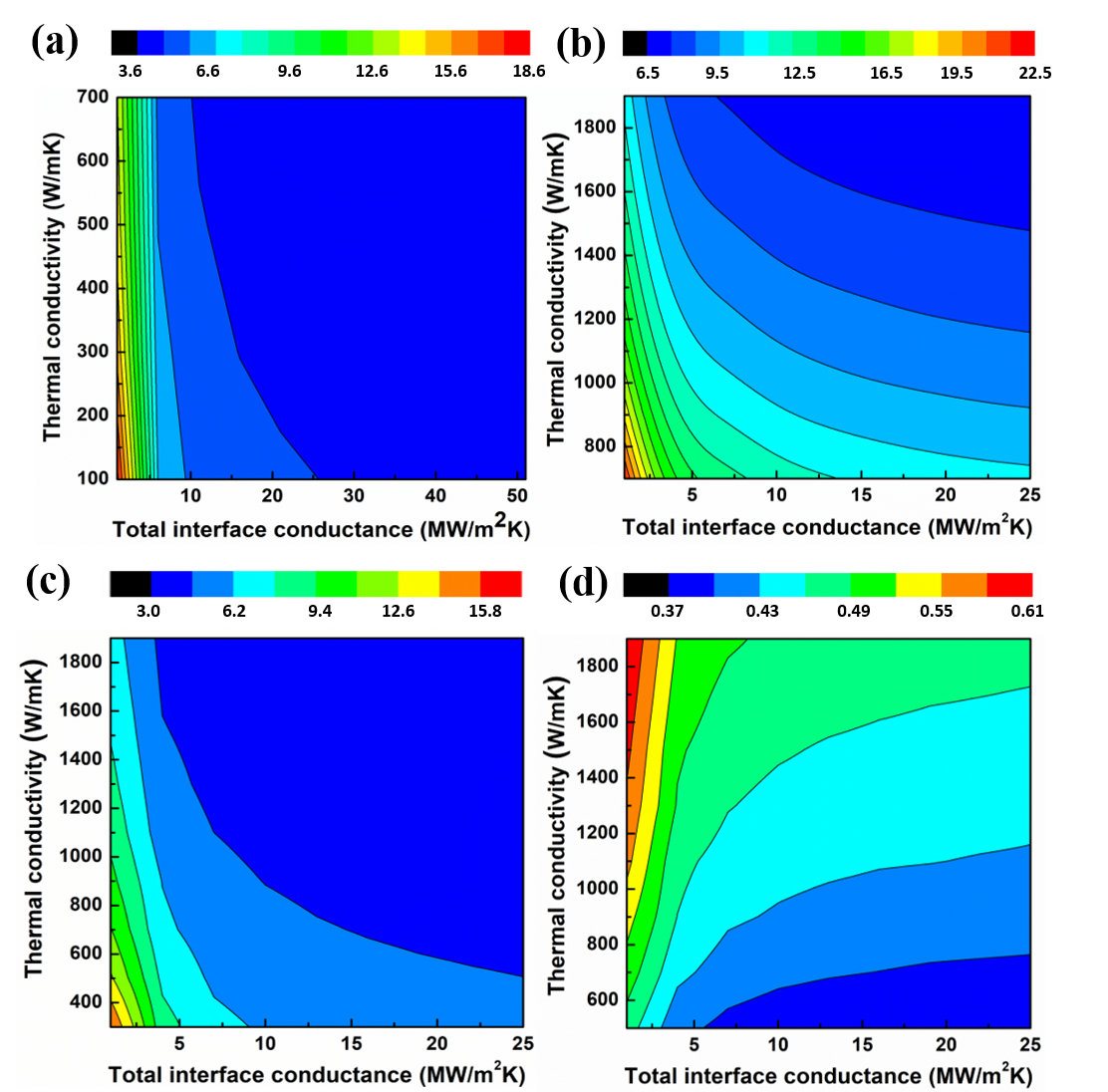}
    \caption{FEA simulation results (a) Temperature rise for turbostratic graphene using $50\times$ objective (b) Temperature rise for AB-stacked graphene using $100\times$ objective (c) Temperature rise for AB-stacked graphene using $50\times$ objective (d) Temperature rise ratio for AB-stacked graphene}
\end{figure}

Figure S\Romannum{6}(a) shows the temperature rise in Kelvin with 50$\times$ objective for turbostratic graphene. Figure S\Romannum{6}(b) shows the temperature rise with 100$\times$ objective for AB-stacked graphene. Figure S\Romannum{6}(c) shows the temperature rise of AB-stacked graphene with 50$\times$ objective. Figure S\Romannum{6}(d) shows the temperature rise ratio T\textsubscript{50$\times$}/T\textsubscript{100$\times$} for AB-stacked graphene.
\pagebreak
\textbf{\flushleft{\large{\Romannum{7}. Detailed analysis of G-Peak FWHM}}}

\renewcommand{\thefigure}{S\Romannum{7}}
\begin{figure}[h]
    \centering
    \includegraphics[width=14.5cm, height=12.5cm]{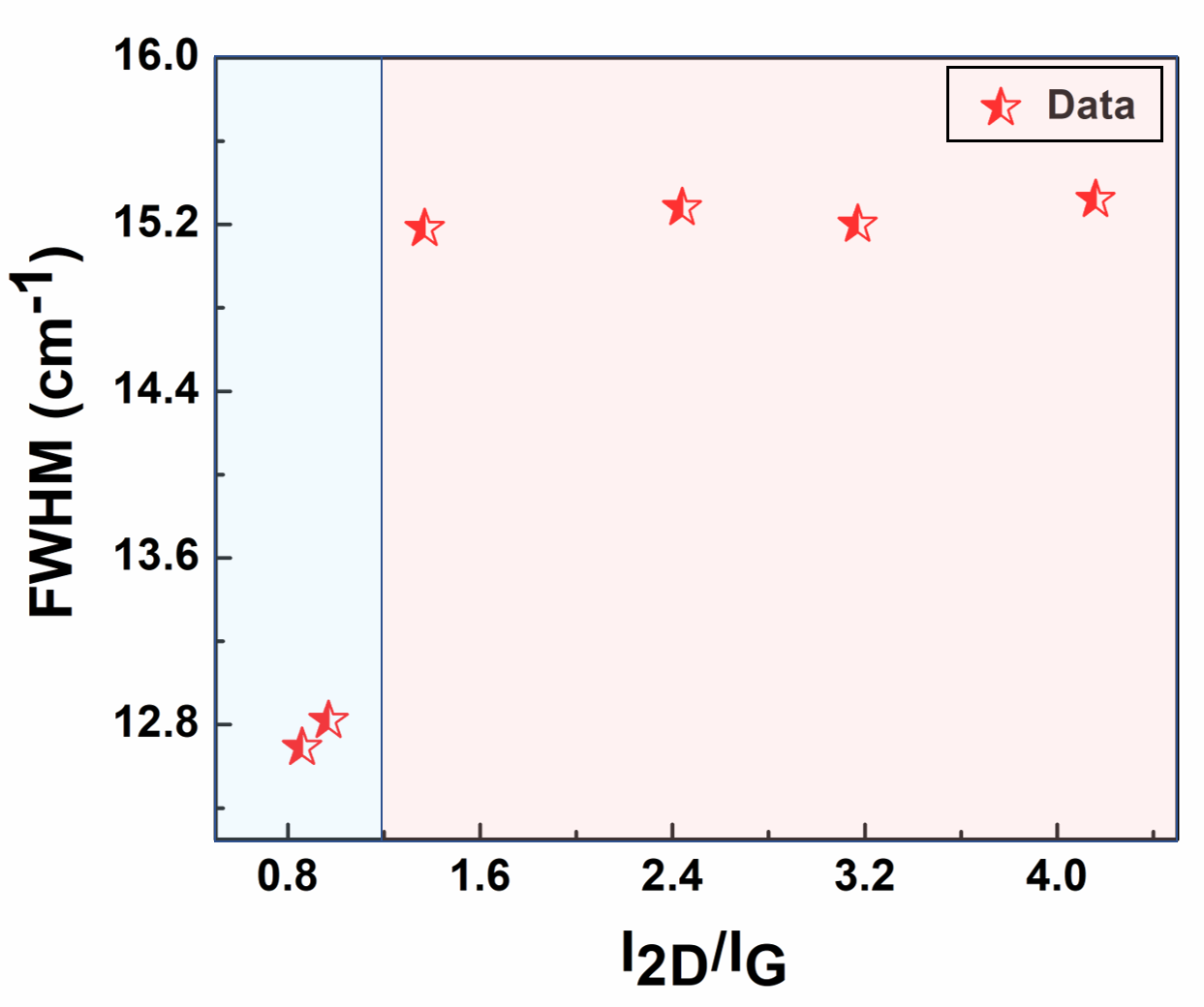}
    \caption{FWHM of AB stacked and turbostratic multilayer graphene as a function of I\textsubscript{2D}/I\textsubscript{G}.}
\end{figure}

Figure S\Romannum{7} shows the FWHM of the AB stacked and turbostratic multilayer graphene as a function of I\textsubscript{2D}/I\textsubscript{G}. Region shaded in cyan color represents the dataset for AB stacked graphene. Region marked in light red shows the dataset for turbostratic graphene. 
\pagebreak
\textbf{\flushleft{\large{\Romannum{8}. Detailed R mode analysis}}}

\renewcommand{\thefigure}{S\Romannum{8}}
\begin{figure}[h]
    \centering
    \includegraphics[width=11cm, height=13cm]{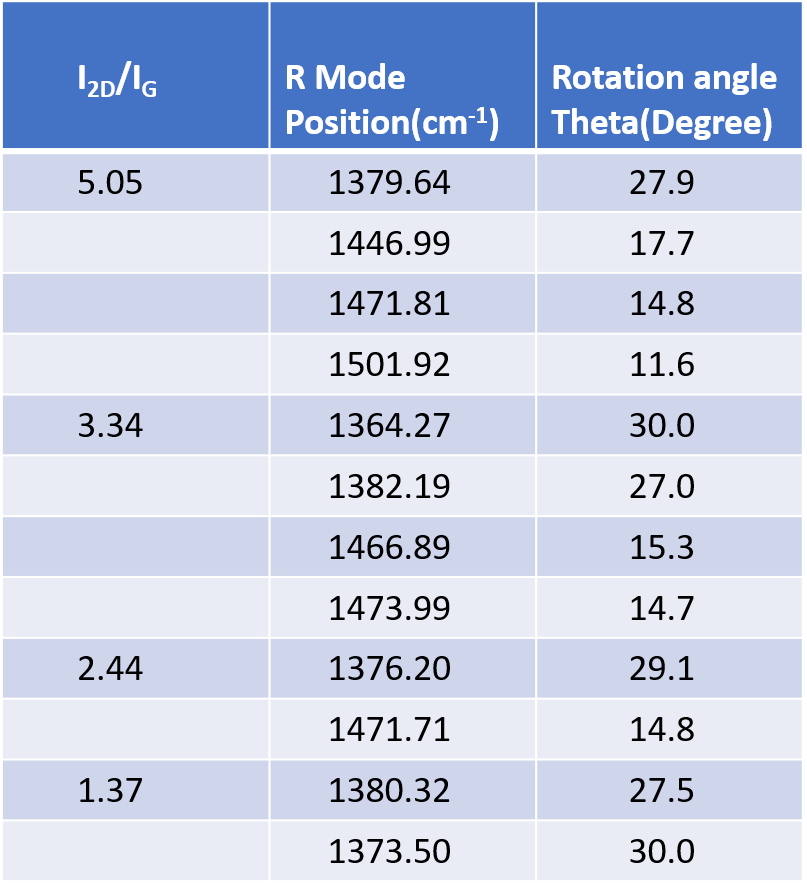}
    \caption{Table lists all the R-mode positions in cm\textsuperscript{-1} with corresponding angles of twist(theta in degree) obtained for samples with different I\textsubscript{2D}/I\textsubscript{G} values.}
\end{figure}

Table lists all the R mode positions obtained by careful acquisition of Raman spectra from samples with different I\textsubscript{2D}/I\textsubscript{G} ratio values.   

\pagebreak

\end{document}